# Microcanonical Treatment of Hadronizing the Quark-Gluon Plasma

Klaus WERNER [§‡]
*Institut für Theoretische Physik, Universität Heidelberg, Germany*

Jörg AICHELIN [*]
*Laboratoire de Physique Subatomique et des Technologies Associées,
Université de Nantes – EMN – IN2P3/CNRS, Nantes, France*

We recently introduced a completely new way to study ultrarelativistic nuclear scattering by providing a link between the string model approach and a statistical description. A key issue is the microcanonical treatment of hadronizing individual quark matter droplets. In this paper we describe in detail the hadronization of these droplets according to n-body phase space, by using methods of statistical physics, i.e. constructing Markov chains of hadron configurations.

## I. INTRODUCTION

Studying nuclear collisions at ultrarelativistic energies ($E_{\mathrm{cms}}/\mathrm{nucleon} \gg 1 \,\mathrm{GeV}$) is motivated mainly by the expectation that a thermalized system of quarks and gluons (quark–gluon plasma) is created [1]. There are essentially two directions for modelling such interactions: dynamical and thermal approaches. The former ones refer to string models [2–7] or related methods [8], supplemented by semihard interactions at very high energies [9–12]. Here, a well established treatment of hadron-hadron scattering, based on Pomerons and AGK rules [13], is extended to nuclear interactions. Thermal methods [14–19] amount to assuming thermalization after some initial time $\tau_0$, with evolution and hadronization being mostly based on ideal gas assumptions.

We recently introduced a completely new approach [20,21], more realistic than the string model and more realistic than thermal approaches, providing a link between the two. Based on the string model, we first determine connected regions of high energy density. These regions are referred to as quark matter (QM) droplets. Presently, a purely longitudinal expansion of the QM droplets is assumed. Once the energy density falls beyond some critical energy density $\varepsilon_c$, the droplet $D$ hadronizes into an $n$-hadron configuration $K = \{h_1 h_2 \ldots h_n\}$ with a probability proportional to $\Omega$, where $\Omega$ represents the microcanonical partition function of an $n$-hadron system. Due to the huge configuration space, sophisticated methods of statistical physics [22,23] have to be employed to solve the problem without further approximations.

So our approach amounts to treating high density regions (droplets) for some time (between formation $\tau_f$ and hadronization $\tau_h$) macroscopically, whereas before $\tau_f$ and after $\tau_h$ a microscopic treatment is employed. What happens – microscopically – between $\tau_f$ and $\tau_h$ is not specified, there may be a first or second order transition, just a crossover, or even some nonequilibrium transition. The macroscopic treatment is chosen due to the lack of appropriate transport theories of dense hadronic or quark matter. So at present we parametrize the behaviour of the dense matter in a simple fashion, the time evolution as longitudinal expansion and a hadronization according to n-body phase space. The hadronization of a droplet is not meant to represent a dynamical description of a phase transition, it means that at $\tau_h$ one observes a multihadron system, whatever happened between $\tau_f$ and $\tau_h$. Whether our parametrization is realistic, and what happens microscopically between $\tau_f$ and $\tau_h$ ramains to be investigated by before mentioned theories.

The first stage of our approach is the identification of high energy density regions, based on the string model, which is already discussed elsewhere [20]. Due to the empirically found correlation, $\bar{y} = \zeta$, between the average rapidity $\bar{y}$ of particles and of the space-time rapidity $\zeta$, a hypersurface $\mathcal{H}_\tau$ of constant proper time $\tau$ can be introduced, in the central region simply defined by $t^2 - z^2 = \tau^2$. After having used the string model (VENUS 5.08) to get complete information on hadron trajectories in space and time, we may now, for given $\tau$, determine energy densities on $\mathcal{H}_\tau$ and thus locate high density regions on $\mathcal{H}_\tau$.

High density regions are considered as QM droplets, presently it is assumed that they expand purely longitudinally. Whenever other droplets or hadrons cross its way, the two objects fuse to form a new, more energetic droplet. Due to the expansion, the energy density of a droplet will at some stage drop below $\varepsilon_c$, which causes hadronization, to be described in the following sections.

We consider the concept of QM droplets to be crucial, in particular at SPS energies. It has been shown [21] that at these energies energy density fluctuations are important: one observes intermediate size regions of high density rather than a uniform distribution. Typical sizes of few tens of fm$^3$ are observed for these high density regions.

## II. HADRONIZATION ACCORDING TO N-BODY PHASE SPACE

For the hadronization of QM droplets we employ the following procedure: the probability of a droplet $D$ with invariant mass $E$ and volume $V$ to hadronize into a configuration $K = \{h_1, \ldots, h_n\}$ of hadrons $h_i$ is given as

$$\text{prob}(D \to K) \sim \Omega(K) , \quad (1)$$

with $\Omega(K)$ being the microcanonical partition function of an ideal, relativistic gas of the $n$ hadrons $h_i$. We first have to define a set $\mathcal{S}$ of hadron species; we take $\mathcal{S}$ to contain the pseudoscalar and vector mesons $(\pi, K, \eta, \eta', \rho, K^*, \omega, \phi)$ and the lowest spin-$\frac{1}{2}$ and spin-$\frac{3}{2}$ baryons $(N, \Lambda, \Sigma, \Xi, \Delta, \Sigma^*, \Xi^*, \Omega)$ and the corresponding antibaryons. A configuration is then an arbitrary set $\{h_1, \ldots, h_n\}$ with $h_i \in \mathcal{S}$.

The partition function is given as

$$\Omega(K) = C_{\text{vol}} \, C_{\text{deg}} \, C_{\text{ident}} \, \phi , \quad (2)$$

with

$$C_{\text{vol}} = \frac{V^n}{(2\pi\hbar)^{3n}} , \quad C_{\text{deg}} = \prod_{i=1}^{n} g_i , \quad C_{\text{ident}} = \prod_{\alpha \in \mathcal{S}} \frac{1}{n_\alpha!} . \quad (3)$$

Here, $C_{\text{deg}}$ accounts for degeneracies ($g_i$ is the degeneracy of particle $i$), and $C_{\text{ident}}$ accounts for the occurrence of identical particles in $K$ ($n_\alpha$ is the number of particles of species $\alpha$). The last factor

$$\phi = \phi(E, m_1, \ldots, m_n) \quad (4)$$
$$= \int \prod_{i=1}^{n} d^3 p_i \, \delta(E - \Sigma \varepsilon_i) \, \delta(\Sigma \vec{p}_i) \, \delta_{Q, \Sigma q_i}$$

is the so-called phase space integral, with $\varepsilon_i = \sqrt{m_i^2 + p_i^2}$ being the energy and $\vec{p}_i$ the 3-momentum of particle $i$. The term $\delta_{Q, \Sigma q_i}$ ensures flavour conservation; $q_i$ is the flavour vector of hadron species $i$, and $Q$ is the flavour vector of the droplet (the components of the flavour vectors represent the net quark content for the quark flavours $u$, $d$, …). The expression eq. (4) is valid for the centre-of-mass frame of the droplet $D$.

We are going to employ Monte Carlo techniques, so we have to generate randomly configurations $K$ according to the probability distribution $\Omega(K)$. We want ot develop a method in particular for intermediate size droplets, covering droplet masses from few GeV up to 100 or 1000 GeV. So the method should work for particle numbers $n = |K|$ between 2 and $10^3$, which means, we have to deal with a huge configuration space. Such problems are well known in statistical physics, and the method at hand is to construct a Markov process, specified by an initial configuration $K_0$, and a transition probability matrix $p(K_i \to K_{i+1})$. In generating a sequence $K_0, K_1, K_2, \ldots$, two fundamental issues have to be payed attention at:

- initial transient: starting usually off equilibrium, it takes a number of iterations, $I_{\text{eq}}$, before one reaches equilibrium;

- autocorrelation in equilibrium: even in equilibrium, subsequent configurations, $K_a$ and $K_{a+i}$, are correlated for some range $I_{\text{auto}}$ of $i$.

In general, both $I_{\text{eq}}$ and $I_{\text{auto}}$ should be as small as possible.

We are going to proceed as follows: for a given droplet $D$ with mass $E$ and volume $V$, we start from some initial configuration $K_0$, and generate a sequence $K_0, K_1, \ldots, K_{I_{\text{eq}}}$, with $I_{\text{eq}}$ being sufficiently large to have reached equilibrium (which is defined to be the steady state of the Markov process). If we repeat this procedure many times, getting configurations $K_{I_{\text{eq}}}^{(1)}, K_{I_{\text{eq}}}^{(2)}, \ldots$, these configurations are distributed as $\Omega(K)$. So for our problem, we have only to deal with the initial transient, not with the autocorrelation in equilibrium. We have to find a transition probability $p$ such that it leads to an equilibrium distribution $\Omega(K)$, with the initial transient $I_{\text{eq}}$ being as small as possible.

So our task is twofold: we need to find efficient ways to calculate, for given $K$, the partition function $\Omega(K)$, and we have to find an appropriate transition probability $p(K_a \to K_b)$.

## III. THE PARTITION FUNCTION $\Omega(K)$

The partition function is given as (eq. (2))

$$\Omega(K) = C_{\text{vol}} \, C_{\text{deg}} \, C_{\text{ident}} \, \phi \quad (5)$$

with the phase-space integral $\phi$ (eq. (4)) and some prefactors $C_i$ (eq. (3)). In the following we discuss methods to calculate $\phi$ for an arbitrary number $n$ of particles, starting with $n = 2$.

For 2 particles ($n = 2$), we have

$$\phi(E, m_1, m_2) = \int d^3 p_1 \, d^3 p_1 \, \delta(E - \varepsilon_1 - \varepsilon_2) \, \delta(\vec{p}_1 + \vec{p}_2)$$
$$= 4\pi \int dp \, p^2 \, \delta(E - \sqrt{m_1^2 + p^2} - \sqrt{m_2^2 + p^2}). \quad (6)$$

With $p_0$ representing the root of the argument of the $\delta$-function,

$$p_0 = \frac{1}{2} \left[ E^2 - 2(m_1^2 + m_2^2) + \frac{1}{E^2}(m_1^2 - m_2^2)^2 \right]^{1/2} , \quad (7)$$

we get

$$\phi(E, m_1, m_2) = 4\pi p_0 \left[ (m_1^2 + p_0^2)^{-\frac{1}{2}} + (m_2^2 + p_0^2)^{-\frac{1}{2}} \right]^{-1} . \quad (8)$$

In the following we consider $n \geq 3$. We propose a method to calculate $\phi$, introduced by Hagedorn [24]. The phase-space integral, eq. (4), may be written as

$$\phi(E, m_1, \ldots, m_n) \qquad (9)$$
$$= (4\pi)^n \int \prod_{i=1}^n dp_i \prod_{i=1}^n p_i^2 \, \delta(E - \sum_{i=1}^n \varepsilon_i) \, W(p_1, \ldots, p_n),$$

with $p_i = |\vec{p}_i|$, and with the "random walk function" $W$ given as

$$W(p_1, \ldots, p_n) := \frac{1}{(4\pi)^n} \int \prod_{i=1}^n d\hat{e}_i \, \delta(\sum_{i=1}^n p_i \hat{e}_i), \qquad (10)$$

with $\hat{e}_i = \vec{p}_i / |\vec{p}_i|$, and with

$$d\hat{e}_i = \sin \vartheta_i \, d\vartheta_i \, d\varphi_i \qquad (11)$$

representing the integration over all directions for a given length $|\vec{p}_i|$ of a momentum vector of a particle. The name "random walk function" is due to the fact that $W$ represents the probability to return back to the origin after $n$ "random walks" $p_i \hat{e}_i$ with given step sizes $p_i$.

We first evaluate $W$ for $n = 3$. One may write

$$W(p_1, p_2, p_3) = \int \prod_{i=1}^3 d^3 q_i \left\{ \prod_{i=1}^3 \frac{\delta(a_i - p_i)}{4\pi a_i^2} \right\} \delta(\Sigma \vec{q}_i), \quad (12)$$

with $a_i := |q_i|$. The integration over $d^3 q_3$ may be performed, and one obtains

$$W(p_1, p_2, p_3) = \frac{1}{(4\pi)^3} \frac{1}{p_1^2 p_2^2 p_3^2} \qquad (13)$$
$$\times \int d^3 q_1 \, d^3 q_2 \, \delta(a_1 - p_1) \delta(a_2 - p_2) \delta(|\vec{q}_1 + \vec{q}_2| - p_3).$$

Taking $\vartheta$ to be the angle between $\vec{q}_1$ and $\vec{q}_2$, we have

$$|\vec{q}_1 + \vec{q}_2| = \sqrt{a_1^2 + a_2^2 - 2 a_1 a_2 \cos \vartheta}$$
$$= \sqrt{p_1^2 + p_2^2 - 2 p_1 p_2 \cos \vartheta}, \qquad (14)$$

so we may perform five of the six integrations, to obtain

$$W(p_1, p_2, p_3) = \frac{2(2\pi)^2}{(4\pi)^3} \frac{1}{p_3^2} \qquad (15)$$
$$\times \int_{-1}^1 d\cos \vartheta \, \delta(\sqrt{p_1^2 + p_2^2 - 2 p_1 p_2 \cos \vartheta} - p_3),$$

and we get the final result

$$W(p_1, p_2, p_3) \qquad (16)$$
$$= \begin{cases} (8\pi \, p_1 p_2 p_3)^{-1} & \text{if } |p_1 - p_2| < p_3 < |p_1 + p_2| \\ 0 & \text{else} \end{cases}$$

In the following, we discuss methods to calculate $W$, for $n \geq 4$. The random walk function may be written as

$$W(p_1, \ldots, p_n) = \frac{1}{(4\pi)^n} \frac{1}{(2\pi)^3} \int d^3 \lambda \int \prod_{j=1}^n d\hat{e}_j \, e^{-i\vec{\lambda} \Sigma p_j \hat{e}_j}, \qquad (17)$$

which leads to

$$W(p_1, \ldots, p_n) = \frac{1}{2\pi^2} \int_0^\infty d\lambda \, \lambda^2 \prod_{j=1}^n \frac{\sin p_j \lambda}{p_j \lambda}. \qquad (18)$$

This is easy to evaluate via numerical integration, as long as $n_{\text{eff}}$ is large ($\geq 10$), where $n_{\text{eff}}$ is the number of momenta with $p_i \geq \varepsilon$, with some small $\varepsilon$. Otherwise the integrand fluctuates so strongly, we have to find a different method, as discussed in the following.

From eq. (18), we get

$$W(p_1, \ldots, p_n) = \frac{1}{4\pi^2} \frac{1}{\Pi p_j} \frac{1}{(2i)^n} \qquad (19)$$
$$\times \int_{-\infty - i\varepsilon}^{\infty - i\varepsilon} \frac{d\lambda}{\lambda^{n-2}} \prod_{j=1}^n \left\{ e^{i p_j \lambda} - e^{-i p_j \lambda} \right\}.$$

The product $\prod \{\}$ in eq. (19) can be written as

$$\prod_{j=1}^n \sum_{\sigma_j \in \{1, -1\}} \sigma_j e^{i \lambda \sigma_j p_j}, \qquad (20)$$

which is equal to

$$\sum_{\sigma_1 \ldots \sigma_n} \sigma_1 \ldots \sigma_n \, e^{i \lambda \Sigma \sigma_j p_j}. \qquad (21)$$

For $\Sigma \sigma_j p_j$ being nonnegative, we have

$$\int_{-\infty - i\varepsilon}^{\infty - i\varepsilon} \frac{d\lambda}{\lambda^{n-2}} e^{i \lambda \Sigma \sigma_j p_j}$$
$$= 2\pi i \, \text{Res}\{ \lambda^{2-n} e^{i \lambda \Sigma \sigma_j p_j} \}_{\lambda = 0}$$
$$= 2\pi i \frac{1}{(n-3)!} \left\{ \frac{d^{n-3}}{d\lambda^{n-3}} e^{i \lambda \Sigma \sigma_j p_j} \right\}_{\lambda = 0}$$
$$= 2\pi i \frac{1}{(n-3)!} (i \Sigma \sigma_j p_j)^{n-3}, \qquad (22)$$

whereas for negative $\Sigma \sigma_j p_j$, the integral is zero. So we get

$$W(p_1, \ldots, p_n) = \frac{1}{2^{n+1} \pi \, (n-3)! \, p_1 \ldots p_n} \qquad$$
$$\times \sum_{\substack{\sigma_1 \ldots \sigma_n \\ \Sigma \sigma_j p_j \geq 0}} \sigma_1 \ldots \sigma_n \left( \sum_j \sigma_j p_j \right)^{n-3}. \qquad (23)$$

To perform the summation $\sum_{\sigma_1 \ldots \sigma_n}$, it is useful to take a specific sequence of $\vec{\sigma}$'s (using $\vec{\sigma} = \{\sigma_1, \ldots, \sigma_n\}$), namely [24]

$$\vec{\sigma}^{(1)} = \{+ + + + ...\}$$
$$\vec{\sigma}^{(2)} = \{- + + + ...\}$$
$$\vec{\sigma}^{(3)} = \{- - + + ...\}$$
$$\vec{\sigma}^{(4)} = \{+ - + + ...\}$$
$$\ldots \quad , \quad (24)$$

where we simply write $+$ and $-$ rather than $+1$ and $-1$. The general rule is that $\vec{\sigma}^{(\nu+1)}$ is obtained from $\vec{\sigma}^{(\nu)}$, by changing position number $j_\nu$, where $j_1, j_2, j_3 \ldots$ is given as

$$1, 2, 1, 3, 1, 2, 1, 4, 1, 2, 1, \ldots , \quad (25)$$

with obvious continuation; the rule to obtain the sequence $\{j_\nu\}$ is: taking the binary representation of $\nu$, one obtains $j_\nu$ as the position of the right-most non-zero digit, counting from right to left. For example for $\nu = 7 = 111$, we have $j_7 = 1$, for $\nu = 8 = 1000$, we have $j_8 = 4$. With this prescription actually all possible $\vec{\sigma}$'s are accounted for. Using this sequence of $\vec{\sigma}$'s as described above, we obtain

$$\left\{ \sum_{j=1}^{n} \sigma_j p_j \right\}_{\vec{\sigma}^{(\nu+1)}} = \left\{ \sum_{j=1}^{n} \sigma_j p_j \right\}_{\vec{\sigma}^{(\nu)}} - 2\sigma_{j_\nu}^{(\nu)} p_{j_\nu}, \quad (26)$$

which means that rather than performing the whole sum $\sum \sigma_i p_i$, only one term, $-2\sigma_{j_\nu}^{(\nu)} p_{j_\nu}$, has to be added.

Eq. (23) provides a method to calculate $W$, as long as the number $n$ of particles is not too large ($n \leq 20$). We discussed earlier (eq. (18)) a way to calculate $W$ for large $n$ ($n_{\text{eff}} \geq 10$). Fortunately, there is some overlap between the two methods, so we may always use one or the other procedure to evaluate $W$ to any given accuracy. In practice, we use eq. (23) for $n \leq n_0$ (with $n_0 \simeq 10$), for larger $n$ we use eq. (18). It may happen that for $n > n_0$ the desired accuracy cannot be achieved, due to the fact that one or several momenta $p_i$ are small leading to a strong oscillation of the integrand of eq. (18). In this case, we use the other method, eq. (23).

Having a reliable and efficient method to calculate $W$, we may return to the problem of how to calculate the phase space integral $\phi$ efficiently. From eq. (9), we obtain

$$\phi(E, m_1, \ldots, m_n) = (4\pi)^n \int_{m_1}^{\infty} d\varepsilon_1 \ldots \int_{m_n}^{\infty} d\varepsilon_n \prod_{i=1}^{n} p_i \varepsilon_i$$
$$\times \delta(E - \sum_{i=1}^{n} \varepsilon_i) W(p_1, \ldots, p_n), \quad (27)$$

after a change of variables toward particle energies $\varepsilon_i = \sqrt{m_i^2 + p_i^2}$. We introduce kinetic energy variables,

$$t_i := \varepsilon_i - m_i, \quad (28)$$

and a total kinetic energy $T$,

$$T := E - \sum_{i=1}^{n} m_i, \quad (29)$$

and obtain

$$\phi(E, m_1, \ldots, m_n) = (4\pi)^n \int_0^\infty dt_1 \ldots \int_0^\infty dt_n \prod_{i=1}^n p_i \varepsilon_i$$
$$\times \delta(T - \sum_{i=1}^n t_i) W(p_1, \ldots, p_n). \quad (30)$$

We now introduce "accumulated" kinetic energies $s_i$ via

$$s_i := \sum_{i=1}^{i} t_j, \quad (31)$$

with the inverse

$$t_i = s_i - s_{i-1}, \quad s_0 = 0. \quad (32)$$

The variables are replaced successively,

$$\begin{aligned} t_1 &= s_1, & dt_1 &= ds_1, \\ t_2 &= s_2 - s_1, & dt_2 &= ds_2, \\ t_n &= s_n - s_{n-1}, & dt_n &= ds_n, \end{aligned} \quad (33)$$

and we obtain

$$\phi(E, m_1, \ldots, m_n) = (4\pi)^n \int_0^\infty ds_1 \int_{s_1}^\infty ds_2 \ldots \int_{s_{n-1}}^\infty ds_n$$
$$\times \prod_{i=1}^n p_i \varepsilon_i \delta(T - s_n) W(p_1, \ldots, p_n). \quad (34)$$

The integration over $s_n$ is trivial and may be performed, to obtain

$$\phi(E, m_1, \ldots, m_n)$$
$$= (4\pi)^n \int_0^\infty ds_1 \int_{s_1}^\infty ds_2 \ldots \int_{s_{n-3}}^\infty ds_{n-2} \int_{s_{n-2}}^T ds_{n-1}$$
$$\times \prod_{i=1}^n p_i \varepsilon_i W(p_1, \ldots, p_n). \quad (35)$$

All upper limits may be replaced by $T$. Introducing energy fractions,

$$x_i := \frac{s_i}{T}, \quad (36)$$

we get

$$\phi(E, m_1, \ldots, m_n) = (4\pi)^n T^{n-1} \quad (37)$$
$$\times \int_{0 \leq x_1 \leq \ldots \leq x_{n-1} \leq 1} dx_1 \ldots dx_{n-1} \prod_{i=1}^n p_i \varepsilon_i W(p_1, \ldots, p_n).$$

Using the definition

$$\psi(p_1, \ldots, p_n) := \frac{(4\pi)^n T^{n-1}}{(n-1)!} \prod_{i=1}^n p_i \varepsilon_i W(p_1, \ldots, p_n), \quad (38)$$

we may write

$$\phi(E, m_1, \ldots, m_n) \qquad (39)$$
$$= (n-1)! \int_{0 \leq x_1 \leq \ldots \leq x_{n-1} \leq 1} dx_1 \ldots dx_{n-1} \psi(x_1, \ldots, x_{n-1}),$$

where $\psi(x_1, \ldots, x_{n-1})$ is meant to be $\psi(p_1, \ldots, p_n)$ with $p_i$ and $\varepsilon_i$ expressed in terms of $x_1, \ldots, x_{n-1}$. This may be solved via Monte Carlo as

$$\phi(E, m_1, \ldots, m_n) = \frac{1}{N} \sum_{\beta=1}^{N} \psi(x_1^{(\beta)} \ldots x_{n-1}^{(\beta)}), \qquad (40)$$

where the $x_i^{(\beta)}$ are ordered random numbers,

$$0 \leq x_1^{(\beta)} \leq x_2^{(\beta)} \leq \ldots \leq x_{n-1}^{(\beta)} \leq 1. \qquad (41)$$

So for each Monte Carlo step, $n-1$ random numbers have to be generated, ordered according to size, and then used to evaluate $\psi(x_1^{(\beta)}, \ldots, x_{n-1}^{(\beta)})$. To avoid ordering, one may introduce the variables

$$z_i := \frac{x_i}{x_{i+1}}, \qquad (42)$$

using the definition $x_n := 1$. We get

$$dx_i = dz_i \, x_{i+1} = dz_i \prod_{j=i+1}^{n-1} z_j, \qquad (43)$$

the last equation holding for $i < n-1$; so we have

$$\prod_{i=1}^{n-1} dx_i = \prod_{i=1}^{n-1} dz_i \prod_{i=1}^{n-2} \prod_{j=i+1}^{n-1} z_j$$
$$= \prod_{i=1}^{n-1} dz_i \prod_{i=1}^{n-1} z_i^{i-1}. \qquad (44)$$

From eq. (39), we get

$$\phi(E, m_1, \ldots, m_n) \qquad (45)$$
$$= \int_0^1 dz_1 \ldots \int_0^1 dz_{n-1} \prod_{i=1}^{n-1} i \, z_i^{i-1} \, \psi(z_1, \ldots, z_{n-1}),$$

where obviously $\psi(z_1, \ldots, z_{n-1})$ is meant to be $\psi(p_1, \ldots, p_n)$ with $p_i$ expressed in terms of $z_i$. We now introduce

$$r_i := \int_0^{z_i} i \, \xi^{i-1} \, d\xi = z_i^i \qquad (46)$$

to obtain

$$\phi(E, m_1, \ldots, m_n) = \int_0^1 dr_1 \ldots \int_0^1 dr_{n-1} \psi(r_1, \ldots, r_{n-1}). \qquad (47)$$

The $r_i$ are now uncorrelated, no ordering is required. A Monte Carlo solution is simply

$$\phi(E, m_1, \ldots, m_n) = \frac{1}{N} \sum_{\beta=1}^{N} \psi(r_1^{(\beta)}, \ldots, r_{n-1}^{(\beta)}), \qquad (48)$$

with uncorrelated random number $r_i^{(\beta)}$. So for each Monte Carlo step $\beta$ the following procedure is followed (we drop the index $\beta$):

- generate $n-1$ random number $r_i$;
- calculate $z_i = \sqrt[i]{r_i}$ and then the energy fractions $x_i = x_{i+1} z_i$, $x_n = 1$;
- calculate the accumulated energies $s_i = T x_i$, and then the kinetic energies $t_i = s_i - s_{i-1}$, using $s_0 = 0$. Then calculate the energies $\varepsilon_i = t_i + m_i$ and the momenta $p_i = \sqrt{t_i(t_i + 2m_i)}$;
- calculate $\psi(p_1, \ldots, p_n)$ according to eq. (38), by using the above methods to calculate $W(p_1, \ldots, p_n)$.

Summing up all the $\psi$'s and dividing by the number $N$ of Monte Carlo iterations (see eq. (48) provides the Monte Carlo result for $\phi(E, m_1, \ldots, m_n)$. Clearly most of the computing time goes into the calculation of $W(p_1, \ldots, p_n)$, in particular into the evaluation of $\sin p_i \lambda / p_i \lambda$ for large $n$ (see eq. (18)).

## IV. THE METROPOLIS ALGORITHM

As mentioned earlier, we want to generate randomly hadron configurations $K = \{h_1, \ldots, h_n\}$ according to the probability distribution $\Omega(K)$, where $\Omega(K)$ is the microcanonical partition function discussed extensively in the previous section. With $K^{(\alpha)}$ being such configurations, mean values of observables $O(K)$ are then simply calculated as

$$< O > = \frac{1}{N} \sum_{\alpha=1}^{N} O(K^{(\alpha)}). \qquad (49)$$

To construct a configuration $K^{(\alpha)}$, for each $\alpha$, a chain of configurations $K_0, K_1, K_2, \ldots, K_{I_{\text{eq}}}$ is constructed, which is characterized by an initial configuration $K_0$ and a random matrix $p(K_i \to K_{i+1})$, which specifies the probability of a configuration $K_i$ being followed by $K_{i+1}$. The number $I_{\text{eq}}$ of iterations must be large enough to ensure equilibrium, only then the randomly generated $K_{I_{\text{eq}}}$ are distributed as $\Omega(K)$, for an appropriate $p$, and we take

$$K^{(\alpha)} = K_{I_{\text{eq}}}. \qquad (50)$$

In the following, we discuss how to construct an "appropriate" $p$, which makes the $K^{(\alpha)}$ being distributed as $\Omega(K)$.

Sufficient for the convergence to $\Omega(K)$ is the detailed balance condition,

$$\Omega(K_a)\, p(K_a \to K_b) = \Omega(K_b)\, p(K_b \to K_a) , \qquad (51)$$

and ergodicity, which means that for any $K_a, K_b$ there must exist some $r$ with the probability to get from $K_a$ to $K_b$ in $r$ steps being nonzero. Henceforth, we use the abbreviations

$$\Omega_a := \Omega(K_a); \quad p_{ab} := p(K_a \to K_b). \qquad (52)$$

Following Metropolis [22], we make the ansatz

$$p_{ab} = w_{ab}\, u_{ab} , \qquad (53)$$

with a so-called proposal matrix $w$ and an acceptance matrix $u$. Detailed balance now reads

$$\frac{u_{ab}}{u_{ba}} = \frac{\Omega_b}{\Omega_a} \frac{w_{ba}}{w_{ab}} , \qquad (54)$$

which is obviously fulfilled for

$$u_{ab} = F\left(\frac{\Omega_b}{\Omega_a} \frac{w_{ba}}{w_{ab}}\right) , \qquad (55)$$

with some function $F$ fulfilling $F(z)/F(z^{-1}) = z$. Following Metropolis [22], we take

$$F(z) = \min(z, 1) . \qquad (56)$$

The power of the method is due to the fact that an arbitrary $w$ may be chosen, in connection with $u$ being given by eq. (55). So the task is twofold: one needs an efficient algorithm to calculate, for given $K$, the weight $\Omega(K)$, and one needs to find an appropriate proposal matrix $w$ which leads to fast convergence (small $I_{\mathrm{eq}}$). The first task can be solved, as shown in the previous section. In the following we discuss about constructing an appropriate matrix $w$.

Most natural, though not necessary, is to consider symmetric proposal matrices, $w_{ab} = w_{ba}$, which simplifies the acceptance matrix to $u_{ab} = F(\Omega_b/\Omega_a)$. This is usually referred to as Metropolis algorithm. Whereas for spin system, it is obvious how to define a symmetric matrix $w$, this is not so clear in our case. We may take spin systems as guidance. A configuration $K$ is per def. a set of hadrons $\{h_1,\ldots,h_n\}$ with the ordering not being relevant, so $\{\pi^0,\pi^0,p\}$ is the same as $\{p,\pi^0,\pi^0\}$. We introduce "microconfigurations" to be sequences $\{h_1,\ldots,h_n\}$ of hadrons, where the ordering does matter. So for a given configuration $K_a = \{h_1,\ldots,h_n\}$ there exist several microconfigurations $\tilde{K}_{aj} = \{h_{\pi_j(1)},\ldots,h_{\pi_j(n)}\}$, with $\pi_j$ representing a permutation. The weight of a microconfiguration is

$$\Omega(\tilde{K}_{aj}) = \frac{1}{n!}\left\{\prod_{\alpha\in\mathcal{S}} n_\alpha!\right\} \Omega(K_a) , \qquad (57)$$

with $n_\alpha$ being the number of hadrons of type $\alpha$. Taking for example $K = \{p,\pi^0,\pi^0\}$, there are three microconfigurations $\{p,\pi^0,\pi^0\}$, $\{\pi^0,p,\pi^0\}$ and $\{\pi^0,\pi^0,p\}$, with weight $\Omega(K)/3$.

So far we deal with sequences $\{h_1,\ldots,h_n\}$ of arbitrary length $n$, to be compared with spin system with fixed lattice size. We therefore introduce "zeros", i.e. we supplement the sequences $\{h_1,\ldots,h_n\}$ by adding $L-n$ zeros, as $\{h_1,\ldots,h_n,0,\ldots,0\}$, to obtain sequences of fixed length $L$. The zeros may be inserted at any place, not necessarily at the end. Therefore the weight of a microconfiguration $K_{aj}$ with zeros relative to the one without, $\tilde{K}_{aj}$, is one divided by the number of possibilities to insert $L-n$ zeros, so from eq. (57) we get

$$\Omega(K_{aj}) = \frac{1}{n!}\left\{\prod_{\alpha\in\mathcal{S}} n_\alpha!\right\} \frac{n!(L-n)!}{L!}\Omega(K_a) . \qquad (58)$$

We now have the analogy with a spin system: we have a one-dimensional lattice of fixed size $L$, with each lattice site containing either a hadron or a zero. Henceforth, we use for microconfigurations with zeros the notation $K_{aj} = \{h_1,\ldots,h_L\}$ with $h_i$ being a hadron or zero.

Since from now on we only consider microconfigurations with zeros ($K_{aj}$) rather than configurations ($K_a$), we are going to write $K_a$ instead of $K_{aj}$, keeping in mind that $a$ represents a double index, and say "configuration" rather than "microconfiguration with zeros". The advantage is that we can use the above formulas specifying the Metropolis algorithm without changes.

We are now in a position to define a symmetric proposal matrix $w(K_a \to K_b)$, with $K_a = \{h_1^a,\ldots,h_L^a\}$ and $K_b = \{h_1^b,\ldots,h_L^b\}$, as

$$w(K_a \to K_b) = \qquad (59)$$
$$\frac{2}{L(L-1)}\sum_{i<j}\left\{\prod_{\substack{k=1\\k\neq i,j}}^{L}\delta_{h_k^a h_k^b}\right\} v(h_i^a h_j^a \to h_i^b h_j^b) ,$$

with

$$v(h_i^a h_j^a \to h_i^b h_j^b) = \begin{cases} |\mathcal{P}(h_i^a h_j^a)|^{-1} & \text{if } h_i^b h_j^b \in \mathcal{P}(h_i^a h_j^a) \\ 0 & \text{else} \end{cases} ,$$
$$\qquad (60)$$

where $\mathcal{P}(h_i^a h_j^a)$ is the set of all pairs $(h_i h_j)$ with the same total flavour as the pair $(h_i^a h_j^a)$. The symbol $|\mathcal{P}|$ refers to the number of pairs of $\mathcal{P}$. The term $\{\}$ in eq. (59) makes sure that up to one pair all hadrons in $K_a$ and $K_b$ are the same, the term $L(L-1)/2$ is the probability to randomly choose some pair of lattice indices $i$ and $j$. So our proposal matrix amounts to randomly choosing a pair in $K_a$, and replacing this pair by some pair with the same flavour, with all possible replacements having the same weight. The proposal matrix is obviously symmetric, since $v$ is symmetric (the symmetry of $v$ is crucial!). We have now fully defined an algorithm, which due to general theorems will converge, but how fast, i.e., how large is $I_{\mathrm{eq}}$? This is going to be investigated later.

## V. A GENERALIZED CONFIGURATION SPACE

So far, a configuration was defined to be given as $K = \{h_1, \ldots, h_n\}$ with $h_i$ specifying hadrons species, for example $K = \{\pi^0, \pi^0, p\}$. The Metropolis Algorithm introduced in the previous section will provide random configuration $K^{(\alpha)}$, which are distributed as $\Omega(K)$. This approach is not yet satisfactory for the following reasons: we do not want to make predictions for multiplicities only, but also consider momentum distributions of the hadrons; the method is also extremely slow due to the fact that, for each Metropolis step, the function $\Omega(K)$ has to be evaluated, which itself requires a Monte Carlo procedure with many iterations. There is a way to cure both problems: one has to consider a generalized configuration space, such that not only hadron species are considered but also hadron momenta.

A naive generalization would be to introduce configurations as $\{h_1, \ldots, h_n; p_1, \ldots, p_n\}$, with $p_i$ representing the particle momenta. The symbols $h_i$ represent again the hadron species. There are two problems about the naive generalization: the momenta are not independent, since their sum must be zero, and, in addition, we loose the symmetry property of the proposal matrix $w(K_a \to K_b)$. This symmetry is not really necessary, but at least one needs to be able to calculate the asymmetry $w(K_a \to K_b)/w(K_b \to K_a)$.

To find a reasonable generalization one should recall the discussion following eq. (27), where a couple of coordinate transformations were applied to calculate the phase space integral $\phi$. The final result was eq. (47),

$$\phi(E, m_1, \ldots, m_n) \qquad (61)$$
$$= \int_0^1 dr_1 \ldots \int_0^1 dr_{n-1} \psi(E, m_1, \ldots, m_n; r_1, \ldots, r_{n-1}),$$

with $\psi$ given in eq. (38) as

$$\psi(E, m_1, \ldots, m_n; r_1, \ldots, r_{n-1}) \qquad (62)$$
$$= \frac{(4\pi)^n T^{n-1}}{(n-1)!} \prod_{i=1}^n p_i \, \varepsilon_i \, W(p_1, \ldots, p_n).$$

Here we also indicate the dependence of $\psi$ on $E$ and $m_1, \ldots, m_n$, which has been dropped in the previous chapter. The symbol $T$ denotes the total kinetic energy $E - \Sigma m_i$, and the absolute values of the momenta are expressed in terms of the $r_i$ as

$$p_i = \sqrt{t_i(t_i + 2m_i)}$$
$$t_i = T(x_i - x_{i-1}), \; x_0 = 0$$
$$x_i = x_{i+1} \sqrt[i]{r_i}, \; x_n = 1. \qquad (63)$$

Contrary to the $p_i$, the $r_i$ are independent of each other. Based on eq. (61), we introduce generalized configurations $G$ as

$$G = \{h_1, \ldots, h_n; r_1, \ldots, r_{n-1}\}, \qquad (64)$$

where the $r_i$ are related to the momenta $p_i$ via eq. (63). The weight of such a configuration is

$$\Omega(G) = C_{\text{vol}} \, C_{\text{deg}} \, C_{\text{ident}} \, \psi \qquad (65)$$

(see eq. (2)), with $\psi$ given in eq. (62). We always use the same symbol $\Omega$ for the different functions $\Omega(x)$, depending whether $x$ is a configuration $K_a$, a microconfiguration $K_{aj}$ or a generalized configuration $G_a$.

In order to define a symmetric proposal matrix $w(K_a \to K_b)$, we introduced in the previous chapter microconfigurations. We proceed similarly for the generalized configurations $G_a$. For a given $G_a = \{h_1, \ldots, h_n; r_1, \ldots, r_{n-1}\}$, one obtains several microconfigurations $G_{aj}$, by introducing $L - n$ zeros, leading to a sequence of hadrons $\{h_1, \ldots, h_L\}$ of fixed length, with $h_i$ being a hadron or zero. The sequence $r_1, \ldots, r_{n-1}$ represent the momenta of the nonzero $h_i$ (the hadrons). We supplement $r_1, \ldots, r_{n-1}$ by $L - n$ numbers $r_n, \ldots, r_{L-1}$, with $0 \le r_i \le 1$. A generalized microconfiguration is thus given as

$$G_{aj} = \{h_1, \ldots, h_L; r_1, \ldots, r_{L-1}\}, \qquad (66)$$

with weight

$$\Omega(G_{aj}) = C_{\text{vol}} \, C_{\text{deg}} \, C_{\text{ident}} \, C_{\text{micro}} \, \psi, \qquad (67)$$

with $\psi = \psi(E, m_1, \ldots, m_n; r_1, \ldots, r_{n-1})$ given in eq. (62), the prefactors $C_{\text{vol}}, C_{\text{deg}}, C_{\text{ident}}$ given in eq. (3), and the other prefactor given as

$$C_{\text{micro}} = \frac{1}{n!} \left\{ \prod_{\alpha \in \mathcal{S}} n_\alpha! \right\} \frac{n!(L-n)!}{L!} \qquad (68)$$

(see eq. (55)). The $r_i$ for $i \ge n$ seem to be obsolete, since only $r_1, \ldots, r_{n-1}$ are needed to calculate the momenta of the $n$ hadrons, however, the numbers are needed to define a symmetric proposal matrix. As for the configuration $K$, also for the generalized ones, we write simply $G_a$ rather than $G_{aj}$ and drop the term "micro".

We are now going to define a symmetric proposal matrix $w(G_a \to G_b)$, with $G_a$ given as

$$G_a = \{h_1^a, \ldots, h_L^a; r_1^a, \ldots, r_{L-1}^a\}, \qquad (69)$$

and $G_b$ correspondingly. We introduce a "species part",

$$K_a = \{h_1^a, \ldots, h_L^a\}, \qquad (70)$$

and a "momentum part",

$$R_a = \{r_1^a, \ldots, r_{L-1}^a\}, \qquad (71)$$

and corresponding definitions for $K_b$ and $R_b$. We may now define a proposal matrix $w$ as

$$w(G_a \to G_b) = w_{\text{spec}}(K_a \to K_b) \, w_{\text{mom}}(R_a \to R_b), \qquad (72)$$

where the "species matrix" $w_{\text{spec}}$ is defined in eq. (16), with $w$ instead of $w_{\text{spec}}$ being used. The "momentum matrix" is defined as

$$w_{\text{mom}}(R_a \to R_b) = \int \prod_{i=1}^{N_{\text{mom}}-1} dR_{c_i} \, w^1_{\text{mom}}(R_a \to R_{c_1})$$
$$\times w^1_{\text{mom}}(R_{c_1} \to R_{c_2}) \ldots$$
$$\times \ldots w^1_{\text{mom}}(R_{c_{\{N_{\text{mom}}-1\}}} \to R_b), \quad (73)$$

with $dR = dr_1 dr_2 \ldots dr_{L-1}$, and with

$$w^1_{\text{mom}}(R_a \to R_b) = \frac{1}{L-1} \sum_i \prod_{\substack{j=1 \\ j \neq i}}^{L-1} \delta(r_j^a - r_j^b). \quad (74)$$

The term $(L-1)^{-1}$ indicates the probability to randomly choose a position $i$ between 1 and $L-1$, the second part of eq. (74) ensures that all $r_j^a$ for $j \neq i$ are not allowed to be changed, however, the number $r_i^a$ is replaced by an arbitrary number $r_i^b \in [0,1]$ with probability one. The following Monte Carlo procedure generates an "updated" $R_b$, starting from $R_a$, according to eq. (74): choose randomly a position $i$ ($1 \leq i \leq L-1$), and then replace $r_i^a$ by some random number $r \in [0,1]$. This provides $R_b$. Eq. (73) simply accounts for repeating the above procedure $N_{\text{mom}}$ times.

$N_{\text{mom}}$ is an important technical parameter of our procedure (in addition to $I_{\text{eq}}$ and $L$), which may be chosen between 1 and $L-1$. Let us consider the weight $\Omega(G)$ for fixed hadron species $h_1, \ldots, h_L$, but varying momentum variables $R = \{r_1, \ldots, r_{L-1}\}$. Out of the huge $R$ phase space (for large $n$) only a very small region contributes with significant weight; for most values of $R$, $\Omega(G)$ is practically zero. So taking $N_{\text{mom}} = L-1$, representing a complete $R$-update, would frequently propose configurations with zero weight, which are rejected with a large probability. So, one may get trapped for a long time. Clearly, this choice of $N_{\text{mom}}$ leads to large equilibration times $I_{\text{eq}}$. The other extreme, $N_{\text{mom}} = 1$, provides updated configuration very close to the original one. Now it takes a long time to test the available phase space. In particular it might easily happen, that one gets trapped in the neighbourhood as a local maximum. We have actually the following situation: for a given number $n$ of hadrons in $G$, we have a local maximum of $\Omega(G)$ at some $G^{(n)}$, with $\Omega$ dropping very fast with $G$ moving away from $G^{(n)}$. The maximum values $\Omega(G^{(n)})$ for neighbouring $n$'s are not so different though. One easily gets trapped around some $G^{(n)}$, even with $n$ being quite far away from the equilibrium value. So $N_{\text{mom}}$ must be chosen large enough to explore the available phase space without getting trapped at a local maximum, but not too large, to avoid exploring extremely unlikely regions.

## VI. AN ASYMMETRIC PROPOSAL MATRIX

Considering particle ratios, like $n_{\pi^0}/n_{\pi^+}$, we find immediately that we have a very slow convergence, so $I_{\text{eq}}$ is too large for the method to be of practical importance. This is obvious, since the proposal matrix $w$ does not act very democraticly: flavourless particles like $\pi^0, \rho^0$ or also zeros are much more frequently proposed than all the rest. This shortcoming can be fixed by defining $w$ such that two pairs are exchanged rather than one, the first pair being replaced by a completely arbitrary pair, the second one by some pair to guarantee flavour conservation.

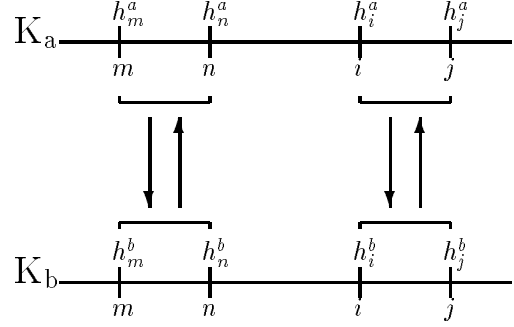

FIG. 1. Double pair exchange.

Such a proposal matrix is not symmetric any more, and the new method is therefore referred to as "asymmetric" or "double-pair-exchange" procedure. In the following, we provide some details about the asymmetric method.

The new proposal matrix is still of the form $w = w_{\text{spec}} w_{\text{mom}}$, see eq. (72), where $w_{\text{spec}}$ refers to particle species and $w_{\text{mom}}$ to momenta. We take the same $w_{\text{mom}}$ as before, only $w_{\text{spec}}$ is changed. We define

$$w_{\text{spec}}(K_a \to K_b) =$$
$$\binom{L}{4}^{-1} \frac{1}{(1+|\mathcal{S}|)^2} \sum_{m<n<i<j} \left\{ \prod_{\substack{k=1 \\ k \neq m,n,i,j}}^{L} \delta_{h_k^a h_k^b} \right\}$$
$$\times v\left(\{h_m^a h_n^a, \bar{h}_m^b, \bar{h}_n^b, h_i^a h_j^a\} \to h_i^b h_j^b\right), \quad (75)$$

with $\bar{h}$ representing the antiparticle of $h$, and with

$$v(\{h_1, h_2, \ldots\} \to h_i^b h_j^b) \qquad\qquad (76)$$
$$= \begin{cases} 0 & \text{if } \mathcal{P}(\{h_1, h_2, \ldots\}) \text{ empty} \\ |\mathcal{P}(\{h_1, h_2, \ldots\})|^{-1} & \text{if } h_i^b h_j^b \in \mathcal{P}(\{h_1, h_2, \ldots\}) \\ 0 & \text{else} \end{cases}.$$

$\mathcal{P}(\{h_1, h_2, \ldots\})$ represents the set of all pairs $h_i h_j$ of hadrons with the same flavour as the set of hadrons $\{h_1, h_2, \ldots\}$. So the double pair exchange works as follows (see fig. 1): two pairs $m < n$ and $i < j$ (with $n < i$) are chosen randomly, with equal probability $\binom{L}{4}$ for all possible double pairs. The first pair $h_n^a h_m^a$ is replaced by some arbitrary pair $h_n^b h_m^b$, all possible pairs having equal weight $(1+|\mathcal{S}|)^{-2}$, with $|\mathcal{S}|$ being the number of hadrons in the basic hadron set $\mathcal{S}$ (containing the standard hadrons, but for testing purposes we will later also use

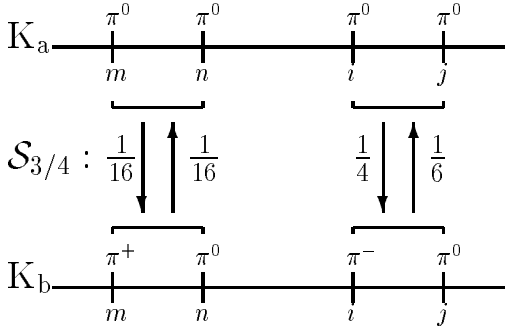

FIG. 2. Example for double pair exchange.

reduced hadron sets). We have $1 + |\mathcal{S}|$ rather than $|\mathcal{S}|$, since we are also considering "zero". To ensure flavour conservation, the second pair $h_i^a h_j^a$ has to be replaced by some pair $h_i^b h_j^b$ having the same flavour as the set of hadrons $\{h_m^a, h_n^a, \bar{h}_m^b, \bar{h}_n^b, h_i^a, h_j^a\}$, where all possible pairs are taken with equal weight.

The new proposal matrix is no more symmetric, which means, the acceptance matrix has to be calculated according to the general expression

$$u(G_a \to G_b) = F\left(\frac{w(G_b \to G_a)}{w(G_a \to G_b)} \frac{\Omega(G_b)}{\Omega(G_a)}\right) \quad (77)$$

rather than using simply $u(G_a \to G_b) = F(\Omega(G_b)/\Omega(G_a))$. Since $w_{\text{mom}}$ is symmetric, the "asymmetry" is given as

$$\frac{w(G_b \to G_a)}{w(G_a \to G_b)} = \frac{w_{\text{spec}}(K_b \to K_a)}{w_{\text{spec}}(K_a \to K_b)}. \quad (78)$$

To evaluate the r.h.s. of eq. (78) we simply need to calculate the ratio of the probabilities $v$ to exchange the second pair, which is given as

$$\frac{|\mathcal{P}(\{h_m^a, h_n^a, \bar{h}_m^b, \bar{h}_n^b, h_i^a, h_j^a\})|^{-1}}{|\mathcal{P}(\{h_m^b, h_n^b, \bar{h}_m^a, \bar{h}_n^a, h_i^b, h_j^b\})|^{-1}}. \quad (79)$$

Let us discuss a simple example (see fig. 2). After choosing randomly positions $m < n < i < j$, we may find for example two pairs each consisting of two $\pi^0$. So we have $h_m^a = h_n^a = h_i^a = h_j^a = \pi^0$. The first pair is replaced by some arbitrary hadron pair, each pair is considered with equal probability $(1+|\mathcal{S}|)^{-2}$. Lets choose a $(\pi^+, \pi^0)$ pair. In order to achieve flavour conservation, the new second pair must have the same flavour as the set of hadrons $\{h_m^a, h_n^a, \bar{h}_m^b, \bar{h}_n^b, h_i^a, h_j^a\}$, which is in this case $\{\pi^0, \pi^0, \bar{\pi}^+, \bar{\pi}^0, \pi^0, \pi^0\}$, so the flavour is $\bar{u}d$. Taking just for testing purposes a reduced hadron set $\mathcal{S}_{3/4} = \{\pi^0, \pi^+, \pi^-\}$, the set of pairs with flavour $\bar{u}d$ is

$$\{(0, \pi^-), (\pi^0, \pi^-), (\pi^-, 0), (\pi^-, \pi^0)\}. \quad (80)$$

Taking equal probabilities, the weight to choose any of these, for example $(\pi^-, \pi^0)$, is $1/4$. Taking the inverse case, we have the two pairs $(h_i^b, h_n^b) = (\pi^+, \pi^0)$ and $(h_i^b, h_j^b) = (\pi^-, \pi^0)$ where the first pair, $(\pi^+, \pi^0)$ is replaced by $(h_m^a, h_n^a) = (\pi^0, \pi^0)$, with the probability for this replacement being $(1 + |\mathcal{S}|)^{-2} = 1/16$. The second pair, $(\pi^-, \pi^0)$, has to be replaced by $(h_i^a, h_j^a) = (\pi^0, \pi^0)$, but the probability for this is not $1/4$. How many pairs would be possible? The pair must have the flavour of the set $\{h_m^b, h_n^b, \bar{h}_m^a, \bar{h}_n^a, h_i^b, h_j^b\}$, which is here $\{\pi^+, \pi^0, \pi^0, \pi^0, \pi^-, \pi^0\}$, so the flavour must be $0$. The set of possible pairs is

$$\{(0,0), (0, \pi^0), (\pi^0, 0), (\pi^0, \pi^0), (\pi^+\pi^-), (\pi^-\pi^+)\}, \quad (81)$$

so the probability to select a pair is $1/6$. The asymmetry $w(G_b \to G_a)/w(G_a \to G_b)$ is therefore $(1/6)/(1/4) = 2/3$.

The example demonstrates that, indeed, the proposal matrix is in general asymmetric, however the asymmetry can be calculated quite easily. For counting the number of possible pairs, one just has to make sure to account for the ordering, for example $(\pi^+, \pi^-)$ and $(\pi^-, \pi^+)$ must be considered as different pairs.

The basic set $\mathcal{S}$ of hadrons has been defined to contain mesons and (anti)baryons from the two lowest multiplets each. For testing purposes we introduce "test sets" $\mathcal{S}_\mu$, for example, we define $\mathcal{S}_1 = \{\pi^0\}$, with the $\pi^0$ being considered massless, we use $\mathcal{S}_2 = \{\pi^0\}$ as well, but considering the correct mass. The complete list $\mathcal{S}_1, \mathcal{S}_2, \ldots, \mathcal{S}_9, \mathcal{S}_{10} \equiv \mathcal{S}$ is given in table I. We consider test sets with massless hadrons, because in this case an analytical treatment is possible, providing useful checks of our Monte Carlo procedures. We will discuss the analytical treatment and detailed comparisons between analytical and Monte Carlo results later. Presently, we are just interested how fast our asymmetric algorithm leads to convergence, depending on the size of the hadron set. We consider a "test droplet" of size $V = 10$ fm$^3$ with mass $E = 10$ GeV, and we apply our hadronization procedure for the test set $\mathcal{S}_1, \mathcal{S}_3, \mathcal{S}_5, \mathcal{S}_7$, and $\mathcal{S}_9$ (so we restrict ourselves to massless hadrons). In fig. 3 we plot the multiplicity $n$ versus the number of iterations, for different sets $\mathcal{S}_\mu$. One clearly observes a fast convergence for small sets, but for $\mathcal{S}_7$ and in particular $\mathcal{S}_9$ (containing the full set of hadrons, just massless), we have a very slow convergence. We discuss in the next section a method to improve that.

TABLE I. The hadron sets $\mathcal{S}_\mu$. Odd $\mu$ implies massless hadrons, for even $\mu$ the correct masses are considered.

| $\mu$ | hadrons in $\mathcal{S}_\mu$ |
|---|---|
| 1,2 | $\pi^0$ |
| 3,4 | $\pi \equiv \pi^0, \pi^+, \pi^-$ |
| 5,6 | $\pi, p, n, \bar{p}, \bar{n}$ |
| 7,8 | $\pi, K, \eta, \eta', p, n, \Sigma, \Xi$ & antibaryons |
| 9,10 | as 7,8 & $\rho, K^*, \omega, \phi, \Delta, \Sigma^*, \Xi^*, \Omega^-$ & antib. |

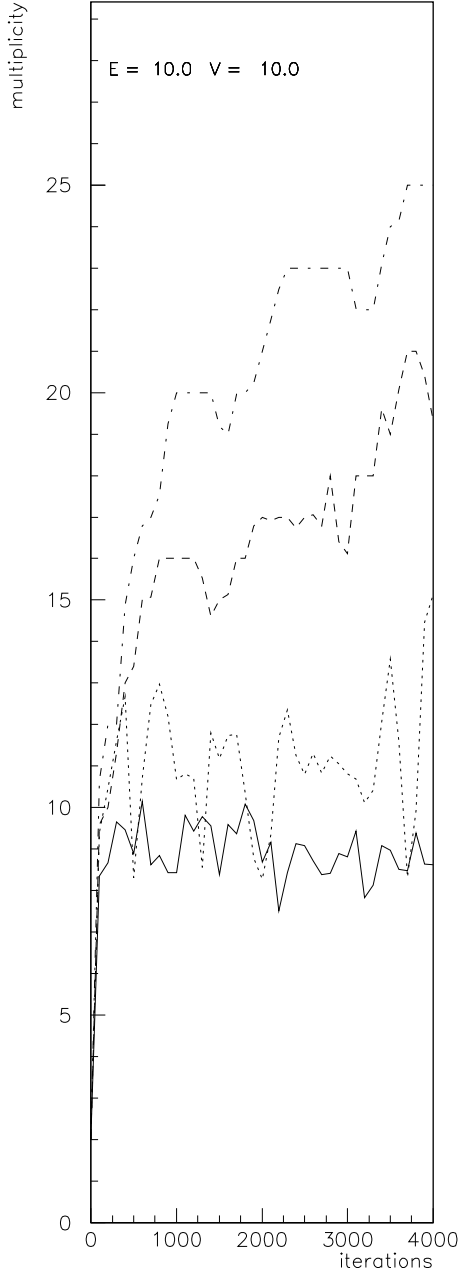

FIG. 3. The multiplicity $n$ versus the number of iterations, for different sets $\mathcal{S}_\mu$: $\mathcal{S}_1$ (solid line), $\mathcal{S}_3$ (dotted), $\mathcal{S}_7$ (dashed), $\mathcal{S}_9$ (dash-dotted). An asymmetric proposal matrix is employed.

## VII. A VERY ASYMMETRIC PROPOSAL MATRIX

So far we have a method which converges fast for small hadron sets ($\mathcal{S}_1, \mathcal{S}_3$) but very slowly for the large ones ($\mathcal{S}_7, \mathcal{S}_9$), and unfortunately the largest is the realistic case. How can one improve the method? We recall that "0" is treated like a hadron: when proposing a new pair $h_1, h_2$, each $h_i$ may be any hadron from $\mathcal{S}_\mu$ or "0", so one may consider extended sets $\tilde{\mathcal{S}}_\mu$, containing the hadrons from $\mathcal{S}_\mu$ and in addition "0". So we have $\tilde{\mathcal{S}}_1 = \{0, \pi^0\}$, $\tilde{\mathcal{S}}_3 = \{0, \pi^0, \pi^+, \pi^-\}$ and so on. A major difference between different $\tilde{\mathcal{S}}_\mu$'s is, that the relative weight of the zeros in $\tilde{\mathcal{S}}_\mu$ decreases with increasing $\mu$: where as for $\tilde{\mathcal{S}}_1$ the zero has weight $1/2$, for $\tilde{\mathcal{S}}_9$ the weight is only $1/55$. On the other hand, large weight implies a large probability to propose a pair containing one or even two zeros, which may reduce the multiplicity, whereas small weight for zeros implies a large probability to propose pairs without zeros, making a reduction of multiplicity rare. So for $\mathcal{S}_9$ or $\mathcal{S}_7$, with small weights for the zeros, there is a large asymmetry in exploring the phase space, it is much more likely to propose a configuration with increased multiplicity than one with reduced multiplicity. Such an asymmetry, causing many unsuccessful suggestions, leads to a slow convergence.

It is obvious how to improve our method: in case of large sets $\mathcal{S}_\mu$, the weight of the "0" must be increased relative to the hadrons. Since in this way we introduce another asymmetry to the proposal matrix $w$, we refer to $w$ as the "very asymmetric" proposal matrix, to distinguish from the "symmetric" case (eq. (59)) and the "asymmetric" case (eq. (75)). The "very asymmetric" proposal matrix $w$ is defined in the following. We again use $w = w_{\text{spec}} w_{\text{mom}}$, see eq. (72), where $w_{\text{mom}}$ is taken as before (eq. (73)), only the "species matrix" $w_{\text{spec}}$ is changed. We define

$$w_{\text{spec}}(K_a \to K_b) = \qquad (82)$$

$$= \binom{L}{4} \frac{1}{(N_{\text{zero}} + |\mathcal{S}|)^2} \sum_{m<n<i<j} \left\{ \prod_{\substack{k=1 \\ k \neq m,n,i,j}}^{L} \delta_{h_k^a h_k^b} \right\}$$

$$\times v\left(\{h_m^a, h_n^a, \bar{h}_m^b, \bar{h}_n^b, h_i^a, h_j^a\} \to h_i^a h_j^a\right) ,$$

with

$$v\left(\{h_1, h_2, \ldots\} \to h_i^b h_j^b\right) = \qquad (83)$$

$$= \begin{cases} 0 & \text{if } \mathcal{P}(\{h_1, h_2, \ldots\}) \text{ empty} \\ \frac{Z(h_i^b) Z(h_j^b)}{\|\mathcal{P}(\{h_1, h_2, \ldots\})\|} & \text{if } h_i^b h_j^b \in \mathcal{P}(\{h_1, h_2, \ldots\}) \\ 0 & \text{else} \end{cases} ,$$

with

$$Z(h) = \begin{cases} N_{\text{zero}} & \text{if } h = 0 \\ 1 & \text{else} \end{cases} . \qquad (84)$$

$\mathcal{P}(\{h_1, h_2, \ldots\})$ represents the set of all pairs $h_i h_j$ of hadrons (or "0") with the same flavour as the set $\{h_1, h_2, \ldots\}$. The symbol $\|\mathcal{P}\|$ represents a weighted sum of pairs of $\mathcal{P}$,

$$\|\mathcal{P}\| = \sum_{h_i h_j \in \mathcal{P}} Z(h_i) Z(h_j). \qquad (85)$$

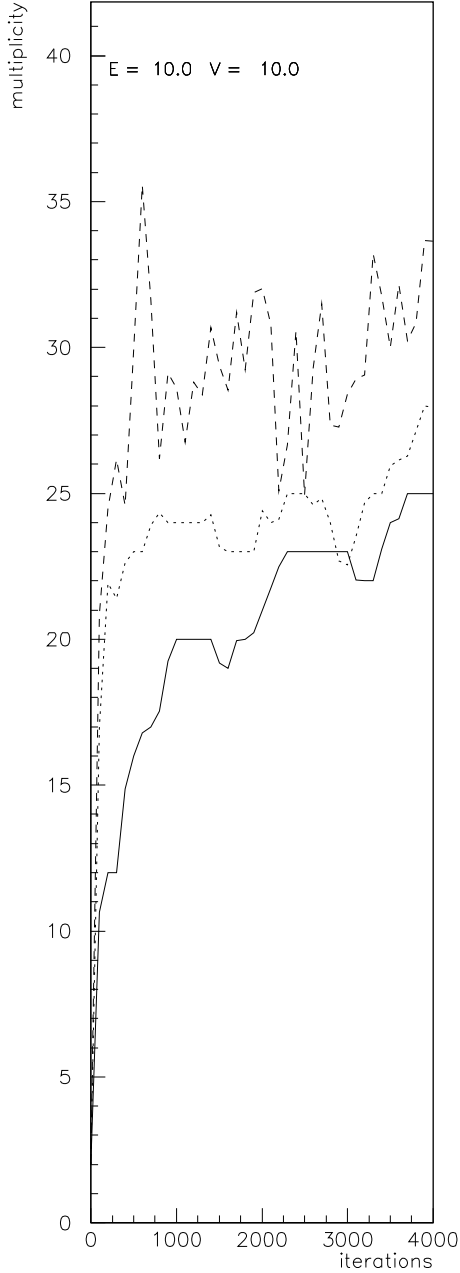

FIG. 4. The multiplicity $n$ versus the number of iterations, for different values of $N_{\text{zero}}$: 1 (solid line), 3 (dotted), 54 (dashed). We refer to a droplet with $E = 10$ GeV, $V = 10$ fm$^3$ and the hadron set $\mathcal{S}_9$.

Whereas in eq. (75) all pairs $h_i^b h_j^b$ have the same relative weight 1, we now consider a relative weight $Z(h_i^b)Z(h_j^b)$, which means, the zero has a weight being $N_{\text{zero}}$ times larger than that of a hadron. Having defined the proposal matrix $w$, we have to determine the asymmetry $w_{\text{spec}}(K_b \to K_a)/w_{\text{spec}}(K_a \to K_b)$, necessary to calculate the acceptance matrix. Similar to eq. (79), the asymmetry is given as

$$\frac{\|\mathcal{P}(\{h_m^a, h_n^a, \bar{h}_m^b, \bar{h}_n^b, h_i^a, h_j^a\})\|^{-1}}{\|\mathcal{P}(\{h_m^b, h_n^b, \bar{h}_m^a, \bar{h}_n^a, h_i^b, h_j^b\})\|^{-1}}, \quad (86)$$

just with $\|\ldots\|$ instead of $|\ldots|$. The "zero weight" $N_{\text{zero}}$ (meant to be an integer larger or equal to one) is a technical parameter, which has to be chosen to guarantee fast convergence for a given hadron set $\mathcal{S}$, in particular for the full set $\mathcal{S}_9$ (or $\mathcal{S}_{10}$).

From the fact that the "asymmetric method" (which corresponds to $N_{\text{zero}} = 1$) works well for $\mathcal{S}_1$, where the weight of the zero is $1/2$, one might expect in general good results for $N_{\text{zero}} = |\mathcal{S}_\mu|$, i.e. $N_{\text{zero}} = 54$ for $\mathcal{S}_9$ or $\mathcal{S}_{10}$. In fig. 4 we see that, indeed, the performance improves significantly by increasing $N_{\text{zero}}$ from 1 to 54.

We did finally set up an algorithm, which seems to be sufficiently fast also for a realistic hadron set ($\mathcal{S}_9$ and $\mathcal{S}_{10}$). In the following sections we are going to compare the Monte Carlo results with analytical calculations.

## VIII. THE ZERO-MASS LIMIT

The hadron masses crucially affect the actual results of the simulations. However, just in order to test the numerical procedures, it is useful to consider the "zero-mass limit", i.e. the case of all hadron masses set equal to zero. In this case analytical results can be obtained, which may be compared with our Monte Carlo simulations. We introduced already, for testing purposes, several basic hadron sets $\mathcal{S}_i$ (see table I), where $\mathcal{S}_1, \mathcal{S}_3, \mathcal{S}_5, \mathcal{S}_7$, and $\mathcal{S}_9$ refer to massless hadrons, with an increasing number of hadrons considered. In the following we discuss the analytical treatment for the case of massless hadrons [25,26].

We consider a configuration with the flavour of the $n$-hadron system, $\sum_{i=1}^n q_i$, being equal to the flavour $Q$ of the droplet $D$. The phase space integral is then

$$\phi = \int \prod_{i=1}^n d^3 p_i \, \delta(E - \Sigma \epsilon_i) \, \delta(\Sigma \vec{p}_i)$$

$$= \frac{d}{dE} \int \prod_{i=1}^n d^3 p_i \, \theta(E - \Sigma \epsilon_i) \, \delta(\Sigma \vec{p}_i). \quad (87)$$

Representing the $\delta$-function as

$$\delta(\vec{x}) = \frac{1}{(2\pi)^3} \int d^3 \lambda \, e^{i \vec{\lambda} \vec{x}} \quad (88)$$

and the $\theta$-function as

$$\theta(x) = \frac{1}{2\pi i} \int_{-\infty - i\epsilon}^{\infty - i\epsilon} \frac{d\alpha}{\alpha} e^{i\alpha x}, \quad (89)$$

the phase space integral may be written as

$$\phi = \frac{1}{i(2\pi)^4} \frac{d}{dE} \int \prod_{i=1}^n d^3 p_i \int d^3 \lambda \, e^{i\vec{\lambda}\Sigma\vec{p}_i} \int \frac{d\alpha}{\alpha} e^{i\alpha(E - \Sigma\epsilon_i)}, \quad (90)$$

which may be rewritten as

$$\phi = \frac{1}{i(2\pi)^4} \frac{d}{dE} \int d\alpha \frac{1}{\alpha} e^{i\alpha E} \int d^3\lambda \prod_{i=1}^{n} I(\lambda, \alpha, m_i), \quad (91)$$

with the "single particle integral" $I_i$ given as

$$I(\alpha, \lambda, m) = \int d^3p \, e^{i\vec{\lambda}\vec{p} - i\alpha\sqrt{m^2+p^2}}. \quad (92)$$

As proven in [25,26] and shown in Appendix A, one may write

$$I(\alpha, \lambda, m) = 2\pi^2 m^2 \frac{\alpha}{\alpha^2 - \lambda^2} H_2^{(2)}(m\sqrt{\alpha^2 - \lambda^2}), \quad (93)$$

with $H_2^{(2)}$ representing a Hankel function. The phase space integral may thus be written as

$$\phi = \phi(E, m_1, \ldots, m_n)$$
$$= \frac{2^{n+1}\pi^{2n}}{(2\pi)^3} \prod_{i=1}^{n} m_i^2 \int_{-\infty-i\varepsilon}^{\infty-i\varepsilon} d\alpha \, \alpha^n e^{i\alpha E}$$
$$\times \int_0^\infty d\lambda \frac{\lambda^2}{(\alpha^2 - \lambda^2)^n} \prod_{i=1}^n H_2^{(2)}(m_i\sqrt{\alpha^2-\lambda^2}) \quad (94)$$

We are now considering the "zero-mass limit", i.e. we calculate

$$\phi_n(E) := \lim_{m_i \to 0} \phi(E, m_1, \ldots, m_n). \quad (95)$$

In this case we may expand the Hankel function about the origin and keep only the first term. We obtain

$$H_2^{(2)}(m_i\sqrt{\alpha^2-\lambda^2}) = \frac{4i}{\pi m_i^2(\alpha^2-\lambda^2)}, \quad (96)$$

which may be substituted into eq. (94), to obtain

$$\phi_n(E) = \frac{2^{3n-3} i^n}{\pi^{3-n}} \int_{-\infty-i\varepsilon}^{\infty-i\varepsilon} d\alpha \, \alpha^n e^{i\alpha E} H_{2n}, \quad (97)$$

with

$$H_{2n} = \int_{-\infty}^{\infty} d\lambda \frac{\lambda^2}{(\alpha^2 - \lambda^2)^{2n}}. \quad (98)$$

As shown in Appendix B, the integration can be done, and one obtains

$$H_{2n} = -\frac{i\pi}{2^{4n-3}} \frac{(4n-4)!}{(2n-1)!(2n-2)!} \frac{1}{\alpha^{4n-3}}. \quad (99)$$

The phase space integral is thus been given as

$$\phi_n(E) = -\frac{i^{n+1}}{\pi^{2-n} 2^n} \frac{(4n-4)!}{(2n-1)!(2n-2)!} \quad (100)$$
$$\times \int_{-\infty-i\varepsilon}^{\infty-i\varepsilon} d\alpha \frac{e^{i\alpha E}}{\alpha^{3n-3}}. \quad (101)$$

By choosing the contour in the upper half-plane, we obtain for the integral

$$2\pi i \frac{(iE)^{3n-4}}{(3n-4)!}, \quad (102)$$

and so the phase space integral, in the zero-mass limit, is given as

$$\phi_n(E) \equiv \lim_{m_i \to 0} \phi(E, m_1, \ldots, m_n)$$
$$= \frac{\pi^{n-1}}{2^{n-1}} \frac{(4n-4)!}{(2n-1)!(2n-2)!(3n-4)!} E^{3n-4} \quad (103)$$

This expression, eq. (103), can be evaluated easily, and is the basis for calculating multiplicity distributions, as discussed in the next section.

## IX. MULTIPLICITY SPECTRA IN THE ZERO-MASS LIMIT

In this section we demonstrate how to calculate multiplicity spectra in the zero-mass limit, and compare the results with the outcome of our Monte Carlo procedure, introduced earlier. This is not only a valuable check of the complicated numerical procedures, but also a very useful tool for optimizing our algorithm [27].

In our statistical treatment, the weight for a hadron configuration $K = \{h_1, \ldots, h_n\}$ is proportional to the partition function $\Omega(K)$; correspondingly, the probability $P_n$ to find $n$ hadrons, is given as

$$P_n = \frac{1}{Z} \sum_{\substack{n_1 \ldots n_s \\ \Sigma n_\nu = n \\ \Sigma n_\nu q_\nu = Q}} \Omega(K_{n_1 \ldots n_s}). \quad (104)$$

Here, $s = |\mathcal{S}|$ is the number of hadrons in the basic hadron set $\mathcal{S} = \{\sigma_1, \ldots, \sigma_s\}$, and $K_{n_1 \ldots n_s}$ is the configuration with $n_\nu$ hadrons of species $\sigma_\nu$. The condition $\Sigma n_\nu q_\nu = Q$ accounts for flavour conservation, $q_\nu$ is the flavour vector of hadron species $\nu$, and $Q$ is the flavour vector of the droplet. In the zero-mass limit, eq. (104) may be written as

$$P_n = \frac{1}{Z} C_{\mathrm{vol}} \phi_n \sum_{\substack{n_1 \ldots n_s \\ \Sigma n_\nu = n \\ \Sigma n_\nu q_\nu = Q}} \prod_{\nu=1}^{s} \frac{g_\nu^{n_\nu}}{n_\nu!}, \quad (105)$$

where eqs. (2,3) have been used. The $g_\nu$ in eq. (105) have a different meaning than the $g_i$ in eq. (3): $g_\nu$ is the degeneracy of hadron species $\sigma_\nu$. $Z$ is a normalization factor. The prefactor $C_{\mathrm{vol}}$ and the phase space integral $\phi_n$ (in the zero-mass limit, see eq. (103)), do not depend on $n_1, \ldots, n_s$, but only on $n$, and therefore appear in front of the summation symbol. We define

$$P_n^0 := \frac{1}{Z} C_{\mathrm{vol}} \frac{1}{n!} \phi_n, \quad (106)$$

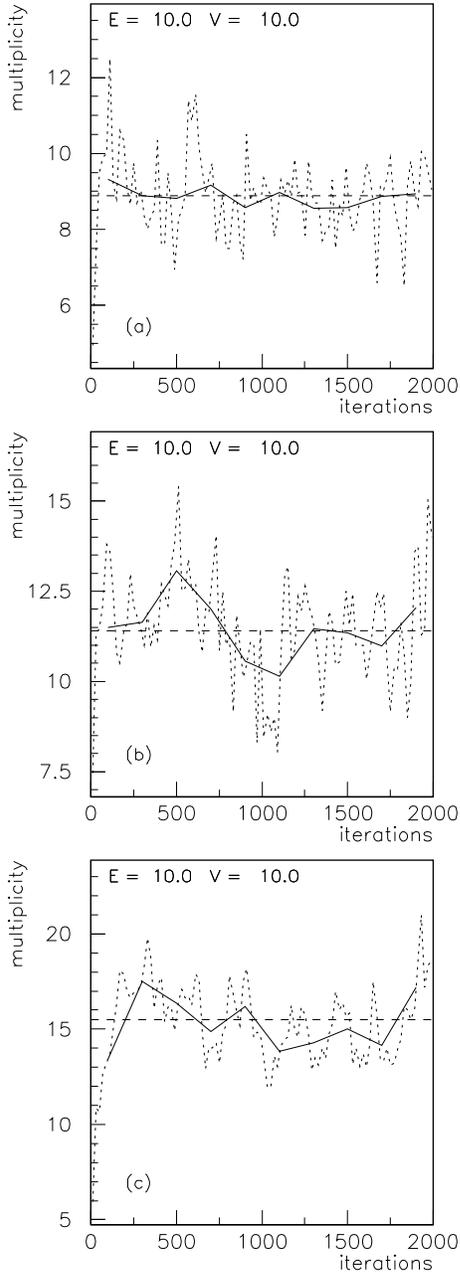

FIG. 5. Medium size droplet: The multiplicity $n$ versus the number of iterations for the hadron sets $\mathcal{S}_1$ (a), $\mathcal{S}_3$ (b), and $\mathcal{S}_5$ (c). The MC results, averaged over 200 iterations (solid lines) and 20 iterations (dotted), are compared with the analytical results for the average multiplicity (dashed).

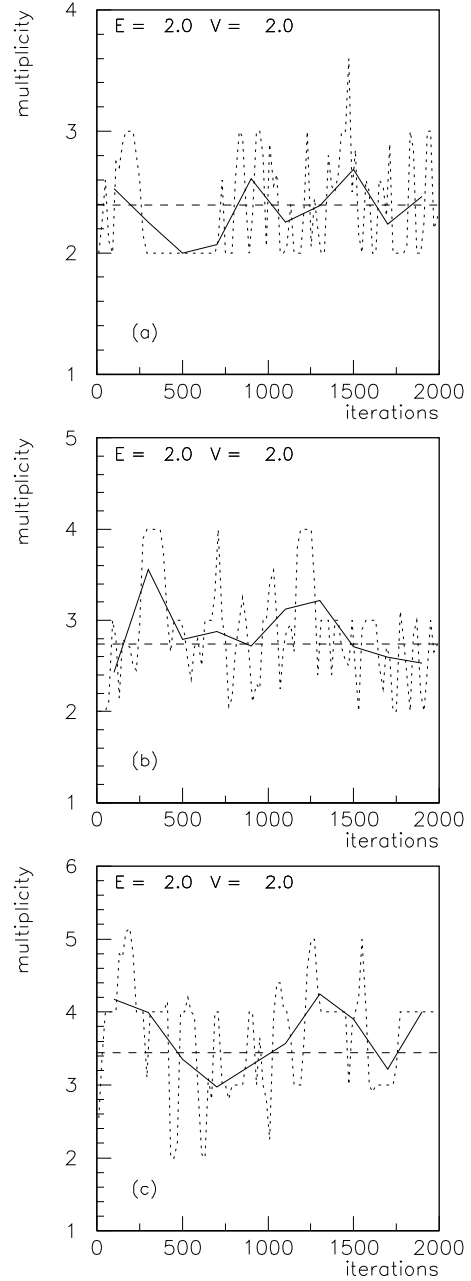

FIG. 6. Small size droplet: The multiplicity $n$ versus the number of iterations for the hadron sets $\mathcal{S}_1$ (a), $\mathcal{S}_3$ (b), and $\mathcal{S}_5$ (c). The MC results, averaged over 200 iterations (solid lines) and 20 iterations (dotted), are compared with the analytical results for the average multiplicity (dashed).

which is equal to $P_n$ for the case of $g_i = 1$ and only one hadron species. In general, we have

$$P_n = P_n^0 \sum_{\substack{n_1 \ldots n_s \\ \Sigma n_\nu = n \\ \Sigma n_\nu q_\nu = Q}} n! \prod_{\nu=1}^{s} \frac{g_\nu^{n_\nu}}{n_\nu!}. \qquad (107)$$

This is the final result for the multiplicity distribution in the zero-mass limit, which can be evaluated numerically as long as $s$ is not too large, for example for our "test sets" $\mathcal{S}_1, \mathcal{S}_3$, or $\mathcal{S}_5$.

In figs. 5 and 6 we compare Monte Carlo (MC) results for the multiplicity with the "analytical results" for the average multiplicity,

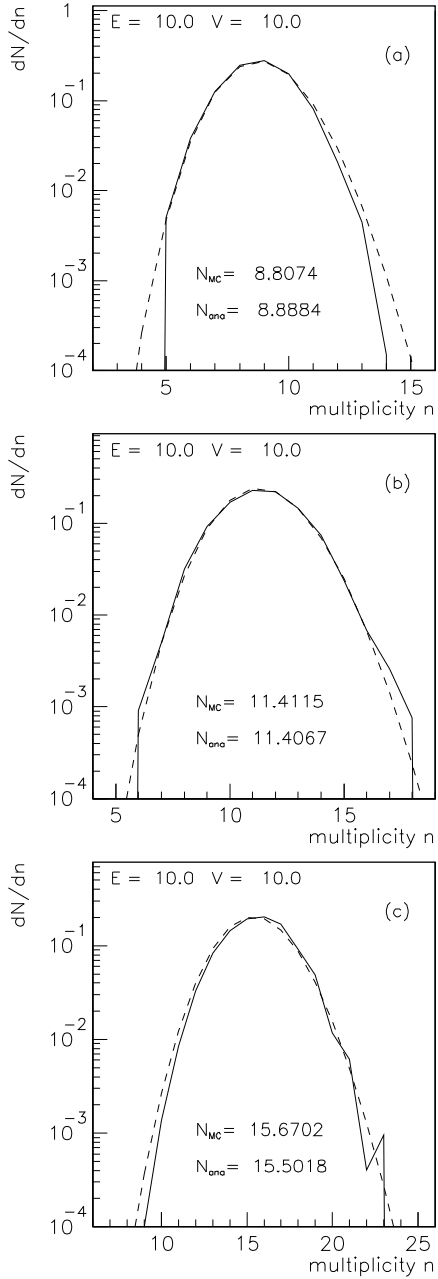

FIG. 7. Medium size droplet: Multiplicity distributions for the hadron sets $\mathcal{S}_1$ (a), $\mathcal{S}_3$ (b), and $\mathcal{S}_5$ (c). MC results (solid lines) are compared with analytical results (dashed).

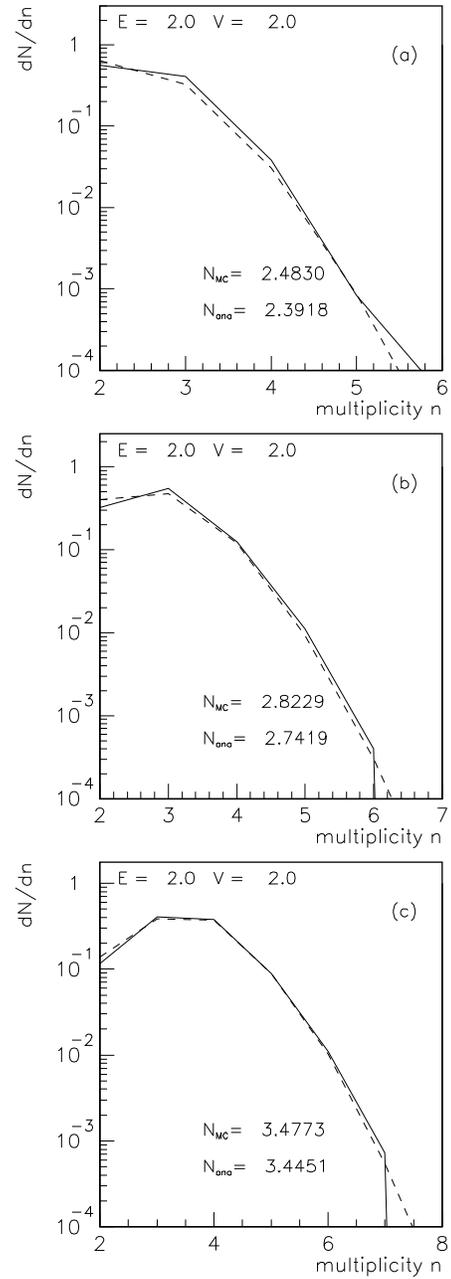

FIG. 8. Small size droplet: Multiplicity distributions for the hadron sets $\mathcal{S}_1$ (a), $\mathcal{S}_3$ (b), and $\mathcal{S}_5$ (c). MC results (solid lines) are compared with analytical results (dashed).

$$<n> = \sum_{n=1}^{\infty} n\, P_n, \qquad (108)$$

with $P_n$ from eq. (107). We consider a medium size droplet ($E = 10$ GeV and $V = 10$ fm$^3$) in fig. 5 and a small size droplet ($E = 2$ GeV and $V = 2$ fm$^3$) in fig. 6; in both cases we have zero net flavour ($Q = 0$). We observe, indeed, that the MC results converge towards the analytical value.

We now turn to multiplicity distributions. In figs. 7 and 8 we compare Monte Carlo (MC) results, again for a medium size droplet ($E = 10$ GeV and $V = 10$ fm$^3$) and for a small size droplet ($E = 2$ GeV and $V = 2$ fm$^3$), with the corresponding "analytical results", obtained from eq. (107). Also the average values $N_{\rm MC}$ and $N_{\rm ana}$ are shown. The MC results are obtained from a single run per spectrum (20000 iterations for the 10 GeV droplet and 200000 for the 2 GeV droplet), which provides an accuracy of about 1 % for the 10 GeV case and of few % for the 2 GeV case for the average multiplicities.

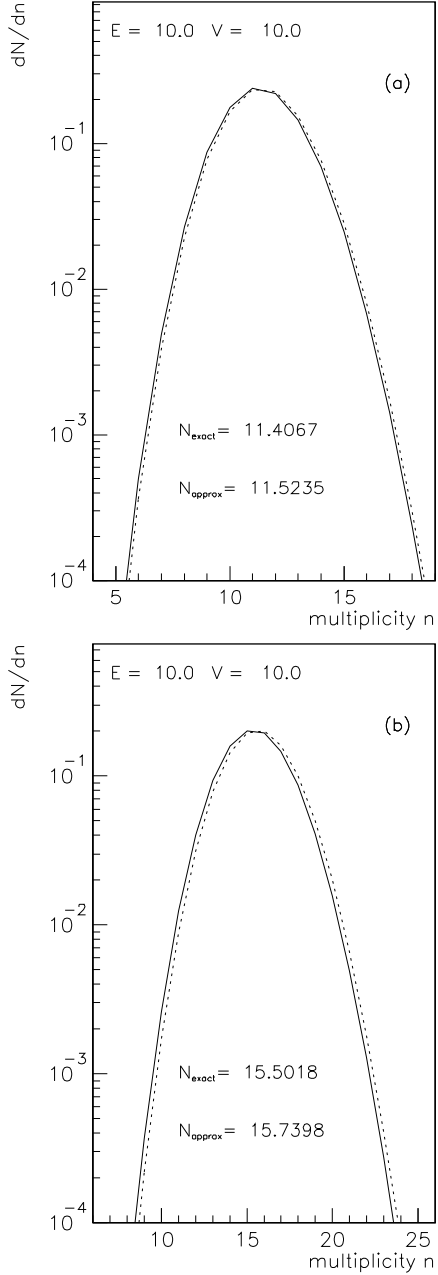

FIG. 9. Medium size droplet: Multiplicity distributions for the hadron sets $\mathcal{S}_3$ (a) and $\mathcal{S}_5$ (b). Exact analytical results (solid lines) are compared with approximate treatments (dotted).

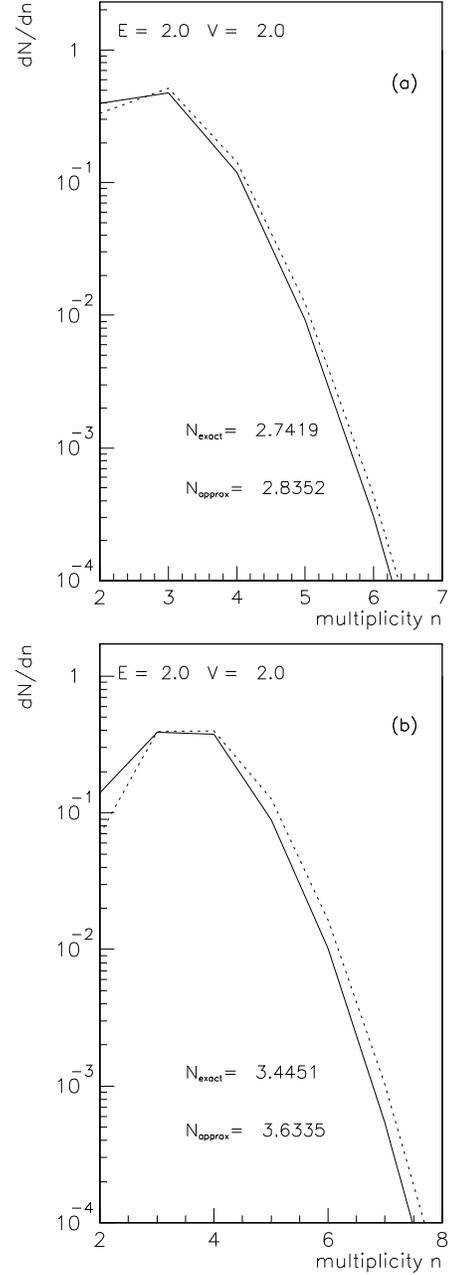

FIG. 10. Small size droplet: Multiplicity distributions for the hadron sets $\mathcal{S}_3$ (a) and $\mathcal{S}_5$ (b). Exact analytical results (solid lines) are compared with approximate treatments (dotted).

For the larger set $\mathcal{S}_7$, and in particular for the realistic set $\mathcal{S}_9$, the exact expression cannot be handled. In this case we use an approximation, by neglecting flavour conservation, i.e., we ignore the condition $\Sigma n_\nu q_\nu = Q$. In this case, using the obvious identity

$$\left(\sum_{\nu=1}^{s} g_\nu\right)^n = \sum_{\substack{n_1 \ldots n_s \\ \Sigma n_\nu = n}} n! \prod_{\nu=1}^{s} \frac{1}{n_\nu!} \prod_{\nu=1}^{s} g_\nu^{n_\nu}, \qquad (109)$$

we get

$$P_n = P_n^0 \left(\sum_{\nu=1}^{s} g_\nu\right)^n, \qquad (110)$$

SUBATECH–95–05        MARCH 1995        HD–TVP–94–24

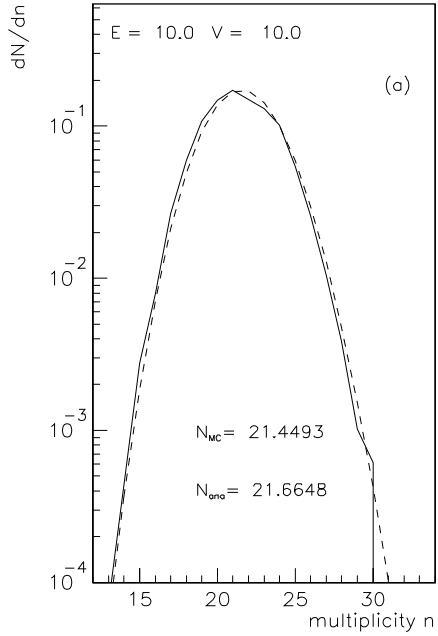
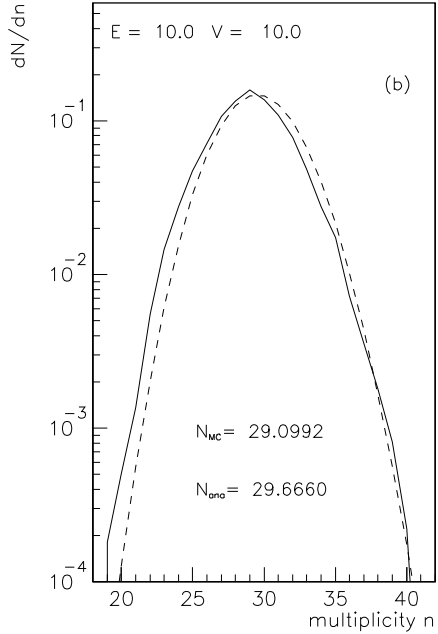

FIG. 11. Medium size droplet: Multiplicity distributions for the hadron sets $\mathcal{S}_7$ (a) and $\mathcal{S}_9$ (b). MC results (solid lines) are compared with the approximate analytical ones (dashed).

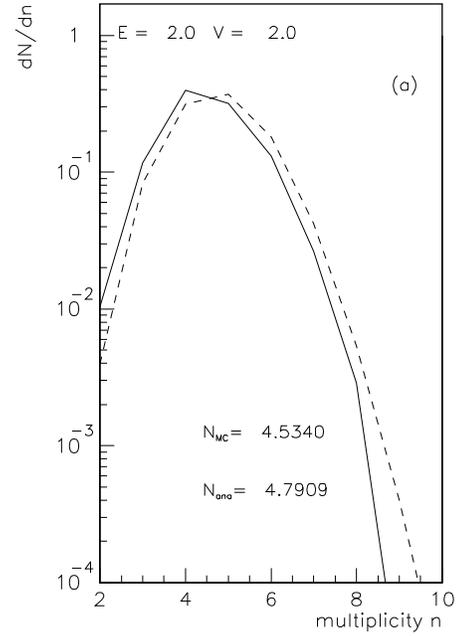
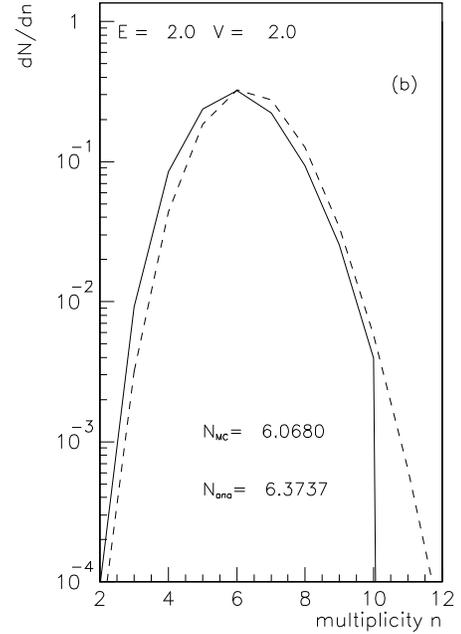

FIG. 12. Small size droplet: Multiplicity distributions for the hadron sets $\mathcal{S}_7$ (a) and $\mathcal{S}_9$ (b). MC results (solid lines) are compared with the approximate analytical ones (dashed).

which can be evaluated also for $\mathcal{S}_7$ and $\mathcal{S}_9$. But first we compare in figs. 9 and 10 the exact and approximate results for the sets $\mathcal{S}_3$ and $\mathcal{S}_5$. Whereas for the medium size droplet the difference is quite small, we observe some disagreement for the small size droplet. We now turn to the large hadron sets: In figs. 11 and 12, we plot multiplicity distributions for the sets $\mathcal{S}_7$ and $\mathcal{S}_9$, comparing MC results with the approximate analytical spectra. The MC spectra are shifted towards somewhat smaller multiplicities, which is consistent with the observation in figs. 9 and 10 that the exact results are "left-shifted" compared to the approximate ones.

The multiplicity distribution $P_n$ refers to "total multiplicities", counting all hadrons. More information provide so-called "partial multiplicities", where only the hadrons of a certain species are counted. We therefore

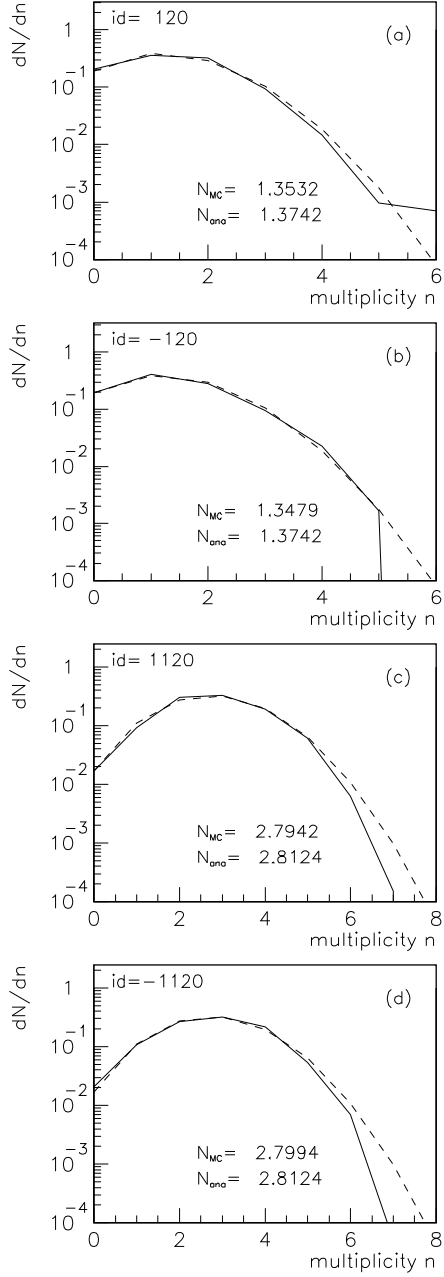

FIG. 13. Medium size droplet: Multiplicity distributions for $\pi^+$ (a), $\pi^-$ (b), $p$ (c), and $\bar{p}$ (d) for the hadron set $\mathcal{S}_5$. MC results (solid lines) are compared with the analytical ones (dashed).

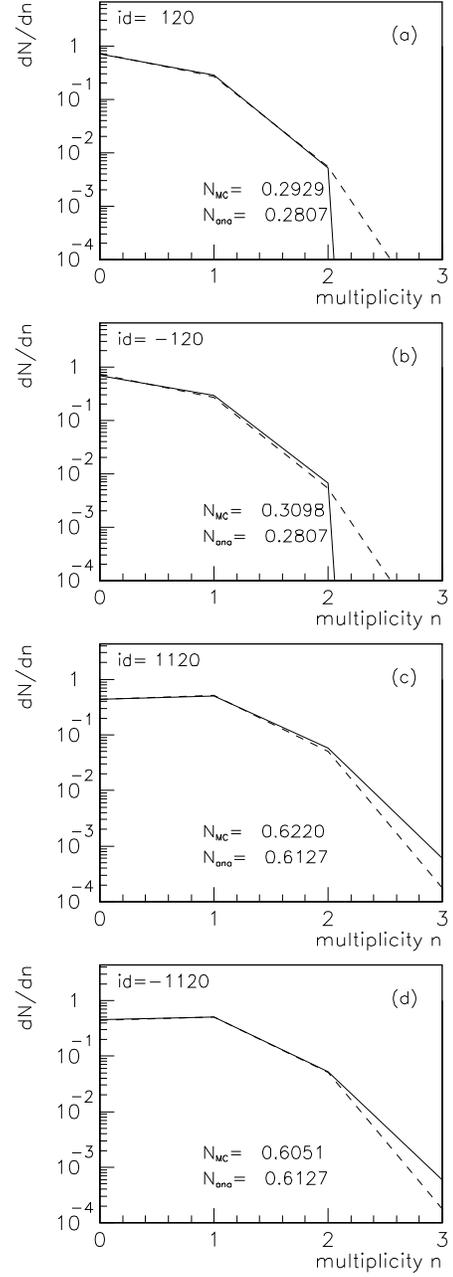

FIG. 14. Small size droplet: Multiplicity distributions for $\pi^+$ (a), $\pi^-$ (b), $p$ (c), and $\bar{p}$ (d) for the hadron set $\mathcal{S}_5$. MC results (solid lines) are compared with the analytical ones (dashed).

introduce the multiplicity distribution of hadron species $\mu$ as

$$P^{\mu}_{n_\mu} = \frac{1}{Z} \sum_{\substack{n_1 \ldots n_{\mu-1} n_{\mu+1} \ldots n_s \\ \Sigma n_\nu q_\nu = Q}} \Omega(K_{n_1 \ldots n_s}), \quad (111)$$

where $n_\mu$ is fixed, and all other multiplicities $n_\nu$ (with $\nu \neq \mu$) are summed over. We may write

$$P^{\mu}_{n_\mu} = \sum_{n=n_\mu}^{\infty} P_n^0 \sum_{\substack{n_1 \ldots n_{\mu-1} n_{\mu+1} \ldots n_s \\ \Sigma n_\nu = n \\ \Sigma n_\nu q_\nu = Q}} n! \prod_{\nu=1}^{s} \frac{g_\nu^{n_\nu}}{n_\nu!}. \quad (112)$$

In order to achieve formal similarities to earlier formulas, we rewrite eq. (112) as

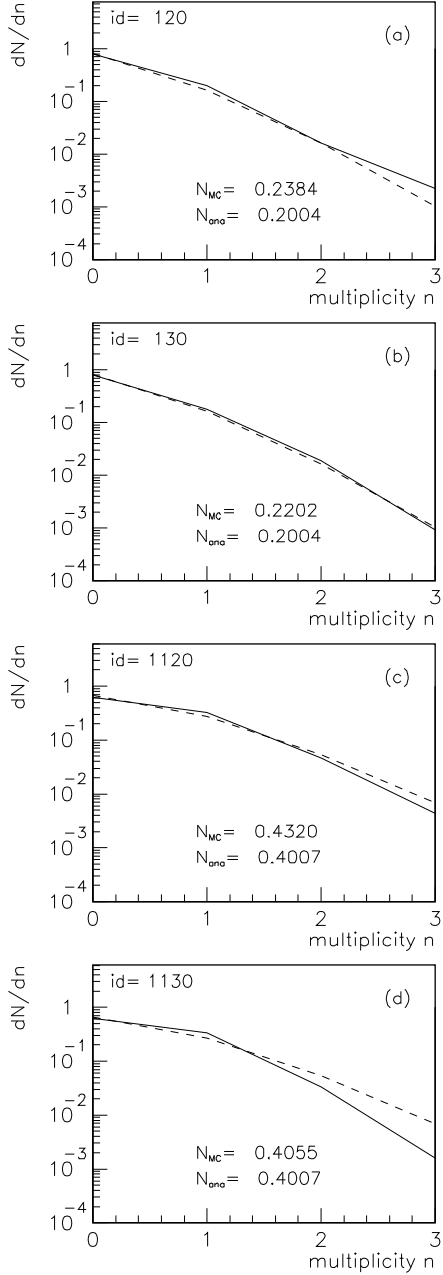

FIG. 15. Medium size droplet: Multiplicity distributions for $\pi^+$ (a), $K^+$ (b), $p$ (c), and $\Lambda$ (d) for the hadron set $\mathcal{S}_9$. MC results (solid lines) are compared with the approximate analytical ones (dashed).

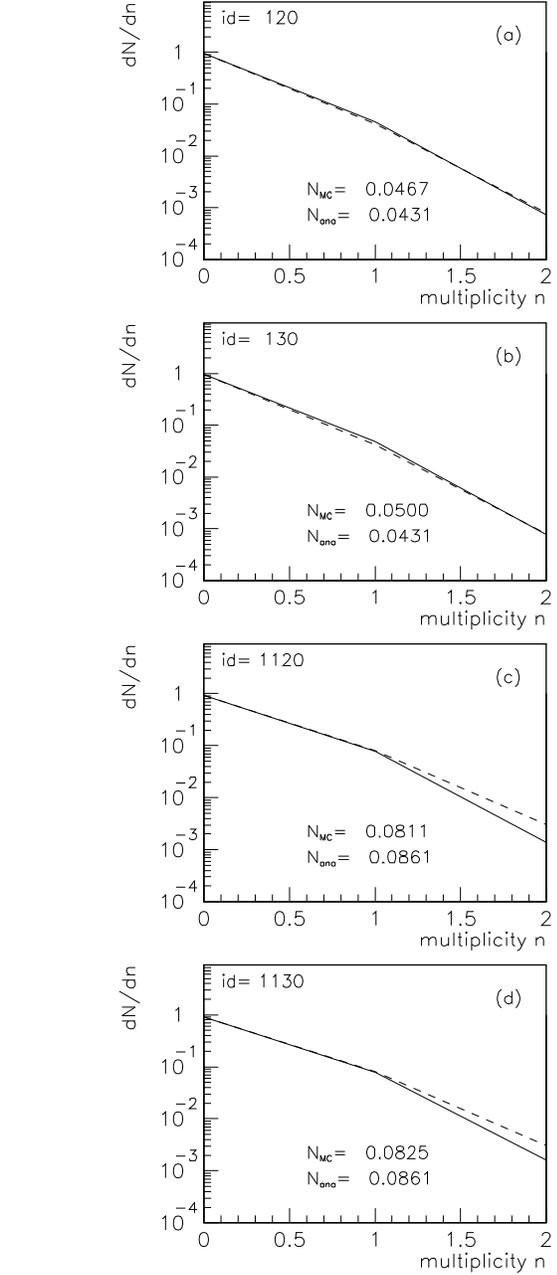

FIG. 16. Small size droplet: Multiplicity distributions for $\pi^+$ (a), $K^+$ (b), $p$ (c), and $\Lambda$ (d) for the hadron set $\mathcal{S}_9$. MC results (solid lines) are compared with the approximate analytical ones (dashed).

$$P^\mu_{n_\mu} = \sum_{n=n_\mu}^\infty \binom{n}{n_\mu} g_\mu^{n_\mu} P_n^0 \qquad (113)$$
$$\times \left\{ \sum_{\substack{n_1\cdots n_{\mu-1} n_{\mu+1}\cdots n_s \\ \Sigma^{(\mu)} n_\nu = n - n_\mu \\ \Sigma^{(\mu)} n_\nu q_\nu = Q - q_\mu}} (n - n_\mu)! \prod_{\nu=1}^n {}^{(\mu)} \frac{g_\nu^{n_\nu}}{n_\nu!} \right\},$$

where $\Sigma^{(\mu)}$ and $\prod^{(\mu)}$ implies looping over $\nu$, except $\nu = \mu$. The expression $\{\}$ in eq. (113) has the same structure as the summation in eq. (107) for total multiplicities, so the same numerical procedures may be employed. In figs. 13 and 14, we show some some multiplicity spectra for specific hadrons. We compare Monte Carlo (MC) results, again for a medium size droplet ($E = 10$ GeV and $V = 10$ fm$^3$) and for a small size droplet ($E = 2$

GeV and $V = 2 \text{ fm}^3$), with the corresponding "analytical results", obtained from eq. (113).

For the larger sets $\mathcal{S}_7$ and $\mathcal{S}_9$, we use the approximation of neclecting flavour conservation. In this case, the simplification eq. (109) may be used, and we obtain

$$P_{n_\mu}^\mu = \sum_{n=n_\mu}^\infty \binom{n}{n_\mu} g_\mu^{n_\mu} P_n^0 \left( \sum_{\nu=1}^s {}^{(\mu)}g_\nu \right)^{n-n_\mu}. \quad (114)$$

In figs. 15 and 16, we show some multiplicity spectra for specific hadrons for the large hadron set $\mathcal{S}_9$. We compare Monte Carlo (MC) results, again for a medium size droplet ($E = 10$ GeV and $V = 10 \text{ fm}^3$) and for a small size droplet ($E = 2$ GeV and $V = 2 \text{ fm}^3$), with the corresponding "approximate analytical results", obtained from eq. (114).

## APPENDIX A

In the following, we prove the relation

$$I \equiv \int d^3p \, e^{i\vec{\lambda}\vec{p} - i\alpha\sqrt{m^2+p^2}}$$
$$= 2\pi^2 m^2 \frac{\alpha}{\alpha^2 - \lambda^2} H_2^{(2)}(m\sqrt{\alpha^2 - \lambda^2}), \quad (A1)$$

with $H_2^{(2)}$ representing a Hankel function, following [25,26].

Introducing polar coordinates, we have

$$I = 2\pi \int_0^\infty dp \, p^2 \int d\cos\vartheta \, e^{i\lambda p \cos\vartheta} e^{-i\alpha\sqrt{m^2+p^2}} \quad (A2)$$

$$= 2\pi \int_0^\infty dp \, p^2 \frac{1}{i\lambda p}\left(e^{i\lambda p} - e^{-i\lambda p}\right) e^{-i\alpha\sqrt{m^2+p^2}} \quad (A3)$$

$$= \frac{2\pi}{i\lambda} \int_{-\infty}^\infty dp \, p \, e^{i\lambda p - i\alpha\sqrt{m^2+p^2}}. \quad (A4)$$

We define $\zeta$, $\lambda_m$, and $\alpha_m$ via

$$p = m\sinh\zeta, \ \lambda_m = \lambda m, \ \alpha_m = \alpha m, \quad (A5)$$

to obtain

$$I = \frac{2\pi}{i\lambda} \int_{-\infty}^\infty d\zeta \, m^2 \cos\zeta \sin\zeta \, e^{i\lambda_m \sinh\zeta - i\alpha_m \cosh\zeta} \quad (A6)$$

$$= -\frac{2\pi m^3}{\lambda_m} \frac{d}{d\lambda_m} \int_{-\infty}^\infty d\zeta \cos\zeta \, e^{i\lambda_m \sinh\zeta - i\alpha_m \cosh\zeta}. \quad (A7)$$

We introduce an auxiliary variable $\varphi$ via

$$\sinh\varphi = \frac{\lambda_m}{\sqrt{\alpha_m^2 - \lambda_m^2}}, \ \cosh\varphi = \frac{\alpha_m}{\sqrt{\alpha_m^2 - \lambda_m^2}}. \quad (A8)$$

The exponent in eq. (A7) may thus be written as

$$i\lambda_m \sinh\zeta - i\alpha_m \cosh\zeta =$$
$$= -i\sqrt{\alpha_m^2 - \lambda_m^2} \quad (A9)$$
$$\times \left(-\sinh\varphi \sinh\zeta + \cosh\varphi \cosh\zeta\right)$$
$$= -i\sqrt{\alpha_m^2 - \lambda_m^2} \cosh(\zeta - \varphi). \quad (A10)$$

With the integration variable

$$\xi := \zeta - \varphi, \quad (A11)$$

using the identity

$$\cosh\zeta = \cosh\xi\cosh\varphi + \sinh\xi\sinh\varphi, \quad (A12)$$

eq. (A7) may be written as

$$I = -\frac{2\pi m^3}{\lambda_m} \frac{d}{d\lambda_m} \frac{\alpha_m}{\sqrt{\alpha_m^2 - \lambda_m^2}}$$
$$\times \int_{-\infty}^\infty d\xi \cosh\xi \, e^{-i\sqrt{\alpha_m^2 + \lambda_m^2}\cosh\xi}. \quad (A13)$$

Introducing

$$z_m := \sqrt{\alpha_m^2 - \lambda_m^2}, \ \frac{d}{d\lambda_m} = -\frac{\lambda_m}{z_m}\frac{d}{dz_m}, \quad (A14)$$

we have

$$I = 2\pi m^3 \alpha_m \frac{1}{z_m} \frac{d}{dz_m} \frac{1}{z_m} \int_{-\infty}^\infty d\xi \cosh\xi \, e^{-iz_m \cosh\xi}. \quad (A15)$$

Using the identities (see [25,26])

$$\int_{-\infty}^\infty d\xi \cosh\xi \, e^{-iz\cosh\xi} = -\pi H_1^{(1)}(ze^{i\pi}) = -\pi H_1^{(2)}(z) \quad (A16)$$

and

$$\frac{1}{z}\frac{d}{dz}\frac{1}{z}H_1^{(2)}(z) = -\frac{1}{z^2}H_2^{(2)}(z), \quad (A17)$$

with $H_j^{(i)}$ representing Hankel functions, we obtain

$$I = 2\pi^2 m^3 \alpha_m \frac{1}{z_m^2} H_2^{(2)}(z_m). \quad (A18)$$

Using the definitions for $z_m$, $\alpha_m$, and $\lambda_m$, we obtain the desired identity, eq. (A1).

## APPENDIX B

We are going to prove the relation

$$H_{2n} \equiv \int_{-\infty}^\infty d\lambda \frac{\lambda^2}{(\alpha^2 - \lambda^2)^{2n}} \quad (B1)$$

$$= -\frac{i\pi}{2^{4n-3}} \frac{(4n-4)!}{(2n-1)!(2n-2)!} \frac{1}{\alpha^{4n-3}}, \quad (B2)$$

for $\alpha = \mathrm{Re}\,\alpha - i\varepsilon$. The integrand may be written as

$$\frac{\lambda^2}{(\alpha^2-\lambda^2)^{2n}} = \alpha^2 \frac{1}{(\alpha^2-\lambda^2)^{2n}} - \frac{1}{(\alpha^2-\lambda^2)^{2n-1}} \quad (B3)$$

and, correspondingly, $H_{2n}$ may be expressed as

$$H_{2n} = \alpha^2 A_{2n} - A_{2n-1}, \quad (B4)$$

with $A_n$ being defined as

$$A_m := \int_{-\infty}^{\infty} d\lambda \frac{1}{(\alpha^2-\lambda^2)^m}. \quad (B5)$$

The integrand has one pole in the upper half-plane, at $\lambda = -\alpha = -\mathrm{Re}\,\alpha + i\varepsilon$, since we are considering the case $\alpha = \mathrm{Re}\,\alpha - i\varepsilon$. So we have

$$A_m = 2\pi i \,\mathrm{Res}\left\{(\alpha^2-\lambda^2)^{-m}\right\}_{\lambda=-\alpha}. \quad (B6)$$

Expanding $(\alpha^2-\lambda^2)^{-m}$ around $\varepsilon = \lambda + \alpha$, we obtain

$$(\alpha^2-\lambda^2)^{-m} = (2\alpha)^{-m}\left(1-\frac{\varepsilon}{2\alpha}\right)^{-m}\varepsilon^{-m} \quad (B7)$$

$$= (2\alpha)^{-m}\varepsilon^{-m} \sum_i \binom{-m}{i}\left(-\frac{\varepsilon}{2\alpha}\right)^i. \quad (B8)$$

We can read off the residue of $(\alpha^2-\lambda^2)^{-m}$ and obtain

$$A_m = 2\pi i \,(2\alpha)^{-m}\binom{-m}{m-1}(-2\alpha)^{1-m} \quad (B9)$$

$$= 2\pi i \,(2\alpha)^{1-2m}(-1)^{1-m} \quad (B10)$$

$$\times \frac{-m(-m-1)\ldots(-2m+2)}{1\,2\,\ldots(m-1)}$$

$$= 2\pi i \,(2\alpha)^{1-2m}\frac{(2m-2)!}{(m-1)!(m-1)!}. \quad (B11)$$

Using eq. (87), we obtain

$$H_{2n} = 2\pi i \left[\alpha^2 \frac{(2\alpha)^{1-4n}(4n-2)!}{(2n-1)!(2n-1)!}\right. \quad (B12)$$

$$\left.-\frac{(2\alpha)^{3-4n}(4n-4)!}{(2n-2)!(2n-2)!}\right] \quad (B13)$$

$$= \frac{-\pi i}{(2\alpha)^{4n-3}}\frac{(4n-4)!}{(2n-1)!(2n-2)!}, \quad (B14)$$

which proves eq. (B1).

§ Heisenberg fellow
‡ Internet: werner@tick.mpi-hd.mpg.de
∗ Internet: aichelin@nanvs2.in2p3.fr

[1] Proc. of "Quark Matter 93", Nucl. Phys. A566 (1994)
[2] K. Werner, Physics Reports 232 (1993) 87–299
[3] A. Capella, U. Sukhatme, Chung-I Tan and J. Tran Thanh Van, Physics Reports 236 (1994) 225
[4] A. Kaidalov, Nucl. Phys. A525 (1991)39c
[5] H. J. Möhring, A. Capella, J. Ranft, J. Tran Thanh Van, C. Merino, Nucl. Phys. A525 (1991) 493c
[6] V. D. Toneev, A. S. Amelin and K. K. Gudima preprint GSI-89-52, 1989
[7] B. Andersen, G. Gustafson and B. Nielsson-Almqvist, Nucl. Phys. B281, 289 (1987)
[8] H. Sorge, H. Stöcker and W. Greiner, Nucl. Phys. A498, 567c (1989)
[9] F. E. Paige, Lecture at "Theoretical Advanced Summer Institute", Boulder, CO, USA, 1989
[10] T. Sjöstrand, M. van Zijl, Phys. Rev. D36 (1987) 2019
[11] X. N. Wang and M. Gyulassy, LBL 31036 (1991), LBL 31159 (1991)
[12] K. Geiger and B. Müller, Nucl. Phys. B369 (1992) 600
[13] V. A. Abramovskiĭ, V. N. Gribov, O. V. Kancheli, Sov. J. Nucl. Phys. 18 (1974) 308
[14] P. Koch, B. Müller, and J. Rafelski, Physics Reports 142 (1986) 167
[15] U. Heinz, Kang. S. Lee, E. Schnedermann, in "Quark Gluon Plasma", ed. R. Hwa, World Scientific, 1990, page 471
[16] J. Zimányi, P. Lévai, B. Lucács, and A. Rácz, in "Particle Production in Highly Excited Matter", ed. H.H. Gutbrod and J. Rafelski, Plenum Press, 1993, page 243
[17] K. Redlich, J. Cleymans, H. Satz, and E. Suhonen, in [1]
[18] H.W. Barz, B.L. Friman, J. Knoll, and H. Schulz, Nucl. Phys. A484 (1988) 661
[19] L.P. Cernai, J.I. Kapusta, G. Kluge, E.E. Zabrodin, Z. Phys. C58 (1993) 453
[20] K. Werner, Phys. Rev. Lett. 73 (1994) 1594
[21] K. Werner and J. Aichelin, preprint HD–TVP–94–23
[22] N. Metropolis, A.W. Rosenbluth, M.N. Rosenbluth, A.H. Teller, and E. Teller, J. Chem. Phys. 21 (1953) 1087
[23] Zhang X.Z., D.D.E. Gross, Xu S.H., and Zheng Y.M., Nucl. Phys. A461 (1987) 668
[24] F. Cerulus and R. Hagedorn, Sup. Nuo. Cim. IX, X (1958) 646
[25] J.V. Lepore and R.N. Stuart, Phys. Rev. 94 (1954) 1724
[26] R.H. Milburn, Rev. Mod. Phys. 27 (1955) 1
[27] M. Hladik and K. Werner, in progress.